\documentclass{aastex63}

\received{September 14, 2020}
\accepted{December 8, 2020}
\submitjournal{AJ}

\shorttitle{Spectroscopic fourth-order coronagraph}
\shortauthors{Matsuo et al.}

\begin{document}

\title{Spectroscopic fourth-order coronagraph for the characterization of terrestrial planets at small angular separations from host stars}

\correspondingauthor{Taro Matsuo}
\email{matsuo@u.phys.nagoya-u.ac.jp}

\author[0000-0001-7694-5885]{Taro Matsuo}
\affiliation{Department of Particle and Astrophysics, Graduate School of Science, Nagoya University \\ Furocho, Chikusa-ku, Nagoya, Aichi 466-8601, Japan}
\affiliation{Department of Earth and Space Science, Osaka University, 1-1 Machikaneyama-cho. Toyonaka 560-0043}
\affiliation{NASA Ames Research Center, Moffett Field, CA 94035, USA}

\author[0000-0003-2690-7092]{Satoshi Itoh}
\affiliation{Institute of Space and Astronautical Science, Japan Aerospace Exploration Agency\\
Yoshinodai 3-1-1, Chuou-ku, Sagamihara, Kanagawa 252-5210, Japan}

\author[0000-0003-2380-8582]{Yuji Ikeda}
\affiliation{Photocoding, 460-102 Iwakura-Nakamachi, Sakyo-ku, Kyoto 606-0025, Japan}



\begin{abstract}
We propose a new approach for high-contrast imaging at the diffraction limit using segmented telescopes in a \replaced{wide}{modest} observation bandwidth. This concept, named "spectroscopic fourth-order coronagraphy", is based on a fourth-order coronagraph with a focal-plane mask that modulates the complex amplitude of the Airy disk along one direction. While coronagraphs applying the complex amplitude mask can achieve the theoretical limit performance for any arbitrary pupils, the focal plane mask severely limits the bandwidth. Here, focusing on the fact that the focal-plane mask modulates the complex amplitude along one direction, we noticed that the mask can be optimized for each spectral element generated by a spectrograph. We combine the fourth-order coronagraph with two spectrographs to produce a stellar spectrum on the focal plane and reconstruct a white pupil on the Lyot stop. Based on the wavefront analysis of an optical design applying an Offner-type imaging spectrograph, we found that the achievable contrast of this concept is $10^{-10}$ at 1.2 - 1.5 times the diffraction limit over the wavelength range of 650 - 750 $nm$ for the entrance pupil of the LUVOIR telescope. Thus, this coronagraph concept could bring new habitable planet candidates not only around G- and K-type stars beyond 20 - 30 $pc$ but also around very nearby M-type stars. This approach potentially promotes the characterization of the atmospheres of nearby terrestrial planets with future on- and off-axis segmented large telescopes.

\end{abstract}



\section{Introduction} \label{sec:intro}
The spectroscopic characterizaion of the atmosphere of an Earth-like planet is the first step toward characterizing the habitability and discovering signs of life on the planet surface \citep[e.g.,][]{DesMarais+2002, Seager+2016, Kaltenegger+2017, Fujii+2018}. Performing spectroscopic analysis in a wide observation bandwidth is crucial for the detection of various molecules in the atmosphere. Although transit spectroscopy has been technically verified \citep[e.g.,][]{Tsiaras+2019, Benneke+2019}, high-contrast imaging spectroscopy, including using a coronagraph in a visible and nulling interferometer in the mid-infrared region, is still challenging; the high-contrast technologies composed of wavefront compensation and coronagraph masks significantly limit the spectral bandpass to 10 $\%$ \citep{Trauger+2012, Cady+2017}\footnote{The state-of-the-art result on the contrast as a function of the bandwidth is described in Section B1 of the Exoplanet Exploration Program 2019 Technology Plan Appendix}. In particular, the wavefront correction in broadband light is challenging because the fresnel propagations of periodic phase and amplitude irregularities from nonpupil optics generate a linear wavelength dependency of phase on the pupil plane, which is different from that of phase compensation with a deformable mirror on the pupil plane \citep{Shaklan+2006}. Furthermore, enlarging the primary mirror is essential for minimizing the nuisance of the zodiacal light on the signal-to-noise ratio of planet detection and observing more distant planetary systems \citep[e.g.,][]{Kasting+2009}. 

Conversely, future large telescopes, such as extremely large telescopes (ELTs) and the large UV optical infrared (LUVOIR) concept, apply on- and off-axis segmented primary mirrors. Achieving extremely high-contrast on the complicated pupil with gaps between mirrors and obscurations of the secondary mirror and spiders is challenging. The methods for mitigating the impact of the gaps on high-contrast imaging have been extensively studied \citep[e.g.,][]{Guyon+2010, Guyon+2014, Mawet+2011a, Pueyo+2013}. For the off-axis segmented telescopes without any obscurations, such as the LUVOIR-B concept, the promising approach involves using a vector vortex coronagraph with the assistance of deformable mirrors and pupil apodization \citep[e.g.,][]{Ruane+2015}. For on-axis telescopes, such as ELTs and the LUVOIR-A concept, the promising coronagraph involves using an apodized pupil Lyot coronagraph with a binary mask \citep[e.g.,][]{N'Diaye+2016}. However, the number of habitable planet candidates that can be observed with future space telescopes is still limited \citep{Stark+2019}. There are mainly two reasons; the inner working angle of the latter approach is largely limited compared to those of the other coronagraphs designed for off-axis telescopes, and construction of large off-axis telescopes, to which the vector vortex mask can be applied, is technically challenging due to the long distance between the primary and secondary mirrors. Therefore, if high-contrast at a small angular separation from the host stars (i.e., 1-2 times the diffraction limit) can be achieved on the on-axis segmented telescopes, the yield of habitable planets will significantly increase.  

\cite{Guyon+2010, Guyon+2014} introduced a complex amplitude mask on the focal plane of an apodized pupil Lyot coronagraph \citep{Soummer+2003}, as well as a phase-induced amplitude apodization \citep{Guyon+2005}. They proved that the new type of coronagraph works at a very small inner working angle for any arbitrary pupil. This complex mask partially transmits light and introduces an $\pi$-phase simultaneously for the destruction of the Airy disk. Contrarily, the effective bandwidth is strictly limited because the size of the complex mask should be proportional to the wavelength. This type of coronagraph is also more sensitive to the telescope pointing jitter and finite stellar disk for larger telescopes because of its second-order sensitivity to low-order \replaced{aberration}{aberrations} \citep[e.g.,][]{Belikov+2018}. 

Based on this background, we propose an approach for achieving fourth-order null at the diffraction limit on the on- and off-axis segmented telescopes in a wide observation bandwidth: we develop the coronagraph proposed by \cite{Itoh+2020} that applies a complex amplitude mask on the focal plane instead of the combination of the pupil apodization with a focal-plane complex amplitude mask. Focusing on the fact that the coronagraphic mask modulates the complex amplitude along one direction of the focal plane, we apply a spectrograph for dispersing the white light along the direction perpendicular to the modulation one and introduce a new mask optimized for the spectrally-resolved Airy disk instead of the original complex amplitude mask. In other words, since the position of the Airy disk changes with the wavelength, the modulation period of the complex amplitude can be optimized for each spectral element. After the complex amplitude mask, a white pupil is reconstructed using another spectrograph with the same optical parameters, and the Lyot stop rejects the stellar light. We name this concept, "spectroscopic fourth-order coronagraphy." Conversely, there are several new problems produced by the spectroscopic coronagraphy concept. We need to investigate how these problems affect the performance of the coronagraph. 

In this paper, we propose the spectroscopic coronagraphy concept. First, in Section \ref{sec:theory}, we present an overview of this concept and evaluate how much it degrades the performance through analytical description. Based on this analytical investigation, in Section \ref{sec:spectroscopic_coronagraph_design}, we propose an optical design of the spectroscopic coronagraph suitably applying an Offner-type imaging spectrograph. We also evaluate the contrast for the proposed optical design with/without alignment errors. In Section \ref{sec:torelance}, we describe our tolerance analysis, considering the factors unconsidered in the previous sections. 

\section{Theory} \label{sec:theory}
In this section, we propose the spectroscopic fourth-order coronagraphy concept for achieving high-contrast over a wide observation bandwidth. This concept combines the coronagraph proposed by \cite{Itoh+2020} with two spectrographs. First, we present an overview of the concept and the new problems generated by applying the spectrographs. Next, we evaluate the impacts of the problems on the performance of the concept, analytically describing the wavefront propagating through the spectroscopic coronagraph. 

\subsection{Overview of Spectroscopic Fourth-order Coronagraphy} \label{subsec:overview}
A complex amplitude mask modulates both the amplitude and phase simultaneously. As discussed in \cite{Guyon+2010}, the Airy disk of the host star can be nulled inside the Lyot stop by adding an $\pi$ phase shift to a part of the Airy disk. Consequently, a small inner working angle of 1 $\frac{\lambda}{D}$ can be achieved. Since the size of the Airy disk is proportional to the wavelength, the size of the region, to which the $\pi$ phase is added, should also be proportional to the wavelength. However, it is technically difficult to manufacture such a complex amplitude mask, and the stellar light cannot be nulled over the wide observation bandwidth; this works only for the monochromatic light. The coronagraph proposed by \cite{Itoh+2020} also has the same characteristics because it applies the complex amplitude mask to the focal plane. Note, however, that the focal-plane mask modulates the complex amplitude along only one direction.

We present an overview of the fourth-order coronagraph with this one-dimensional modulation mask to show the limitation of the observation bandwidth. First, we set the function of the entrance pupil, $P(x,y)$, to $P(x)P(y)$ for simplicity, where $(x,y)$ is the coordinate system of the pupil plane. In this case, the complex amplitude formed on the focal plane, $A_{F}(\alpha,\beta)$, becomes a multiplication of two functions: $A_{F}(\alpha)$ and $A_{F}(\beta)$, where $(\alpha,\beta)$ represents the coordinate system of the focal plane, and $\alpha$ (or $\beta$) is assumed to be parallel to $x$ (or $y$). The complex amplitude is modulated along only one direction of the focal plane, $\alpha$ or $\beta$. Note that it is not necessary to separate the function of the entrance pupil into two functions of an independent variable; the entrance pupil applied by \cite{Itoh+2020} could not be written as $P(x)P(y)$.

Given that the entrance aperture is a square with a size of $D$, the function of the entrance pupil describes 
\begin{eqnarray}
	\label{entrance_pupil}
	P_{1}(x,y) &=&  P_{1}(x)P_{1}(y) \nonumber \\
	&=& (1-P_{s,x}(x))\mathrm{rect}\left(\frac{x}{D}\right)(1-P_{s,y}(y)) \mathrm{rect}\left(\frac{y}{D}\right),
\end{eqnarray}
where the $\mathrm{rect}$ function represents a rectangular function and is defined as
\[
  \mathrm{rect}(x) = \left\{ \begin{array}{ll}
    1 & (|x| < \frac{1}{2}) \\
    0 & (\mathrm{otherwise}),
  \end{array} \right.
\]
and $P_{s,i}(i)$ represents an obscuration along the $i$ axis on the entrance pupil. If there is no obscuration on the entrance pupil, $P_{s,x}(x) = P_{s,y}(y) = 0$. The aperture efficiency of the entrance pupil along the $i$ axis, $\xi_{i}$, is
\begin{equation}
	\label{xi}
	\xi_{i} = \frac{\int di (1-P_{s,i}(i))}{D}.
\end{equation}
In order to null an on-axis source perfectly, $\xi_{i}$ should be constant along the direction perpendicular to the $i$ axis. Note that, although \cite{Itoh+2020} considered off-axis segmented telescopes as the entrance pupil, this one-dimensional complex amplitude mask could be applied to on-axis telescopes such as the LUVOIR-A concept and ELTs; we mask the shadows due to the secondary mirror and supports such that $\xi_{i}$ is constant along one direction of the entrance pupil. We present an example of the entrance pupil optimized for the LUVOIR-A telescope design in Section \ref{performance_on_actual_pupils}, and the throughput is more impacted by the mask than the off-axis telescopes, such as in the LUVOIR-B concept. Assuming that the coronagraph mask optimized for the central wavelength, $\lambda_{c}$, modulates the complex amplitude along the $\alpha$ axis of the focal plane, the mask is written as
\begin{equation}
	\label{M_alpha}
	M(\alpha) = T_{m} \left( 1- m(\alpha) \right), 	
\end{equation}
where $T_{m}$ is the throughput for the off-axis sources, and $m(\alpha)$ shows the modulation function of the complex amplitude. Using the coordinate system of the focal plane normalized by half of the Airy disk's diameter, $(\alpha_{\frac{\lambda}{D}}, \beta_{\frac{\lambda}{D}})$, the modulation function of the complex amplitude is given as
\begin{equation}
	m \left(\alpha_{\frac{\lambda}{D}} \right) =  \frac{w_{0}}{\xi_{x}} \mathrm{sinc}\left \{w_{0}\left(\frac{\lambda}{\lambda_{c}}\right)\pi \alpha_{\frac{\lambda}{D}}\right \},
\end{equation}
where $w_{0}$ is a positive real number, and $\mathrm{sinc}(x)$ is defined as $\frac{\sin(x)}{x}$. We assumed that the complex amplitude is modulated along $\alpha_{\frac{\lambda}{D}}$. The complex amplitude at the exit pupil, on which the Lyot stop is placed, is
\begin{equation}
	\label{amp_lyot}
	A_{L}(x,y,\lambda) = T_{m}P(y) \left\{P(x) - P(x) * \tilde{m}(x)) \right\},
\end{equation} 
where $*$ represents a convolution operator. $\tilde{m}(x)$ is the Fourier conjugate of $m(\alpha_{\frac{\lambda}{D}})$ and is written as
\begin{eqnarray}
	\label{tilde_mask}
	\tilde{m}(x,\lambda) &=& \int\int d\alpha_{\frac{\lambda}{D}} d\beta_{\frac{\lambda}{D}} m(\alpha_{\frac{\lambda}{D}}) \mathrm{e}^{- 2\pi i \left(\alpha_{\frac{\lambda}{D}} \frac{x}{D}  + \beta_{\frac{\lambda}{D}} \frac{y}{D} \right)} \nonumber \\
	&=& \frac{1}{\xi_{x}}\left(\frac{\lambda_{c}}{\lambda} \right) \mathrm{rect} \left( \frac{x}{w_{0}\left( \frac{\lambda}{\lambda_{c}} \right) D}  \right). 
\end{eqnarray}
When $\lambda$ is equal to the central wavelength, $\lambda_{c}$, $\tilde{m}(x,\lambda)$ becomes a rectangular function with an amplitude of $\frac{1}{\xi_{x}}$ and a width of $w_{0}D$. Therefore, when $w_{0} \geq 2$, the on-axis source is completely nulled on the Lyot stop, which is the same as the entrance pupil. However, $A_{L}(x,y,\lambda) \neq 0$ in the other wavelengths because the Fourier conjugate of the mask has a different amplitude from $\frac{1}{\xi}$. The amount of the stellar leak for the $i$-th order coronagraph is expressed as
\begin{equation}
	\label{leak}
	L_{i}(\lambda) = \left( 1- \frac{\lambda_{c}}{\lambda} \right)^{i}.
\end{equation}
As the $i$-th order increases, the stellar leak for the chromatic light decreases. Note that, to suppress the stellar leak down to $10^{-10}$, the wavelength range should be limited to 0.15 and 2.5 $nm$ for the second- and fourth-order coronagraphs with a central wavelength of 700 $nm$, respectively.  

\begin{figure}
	 \centering
	\includegraphics[scale=0.1,height=7cm,clip]{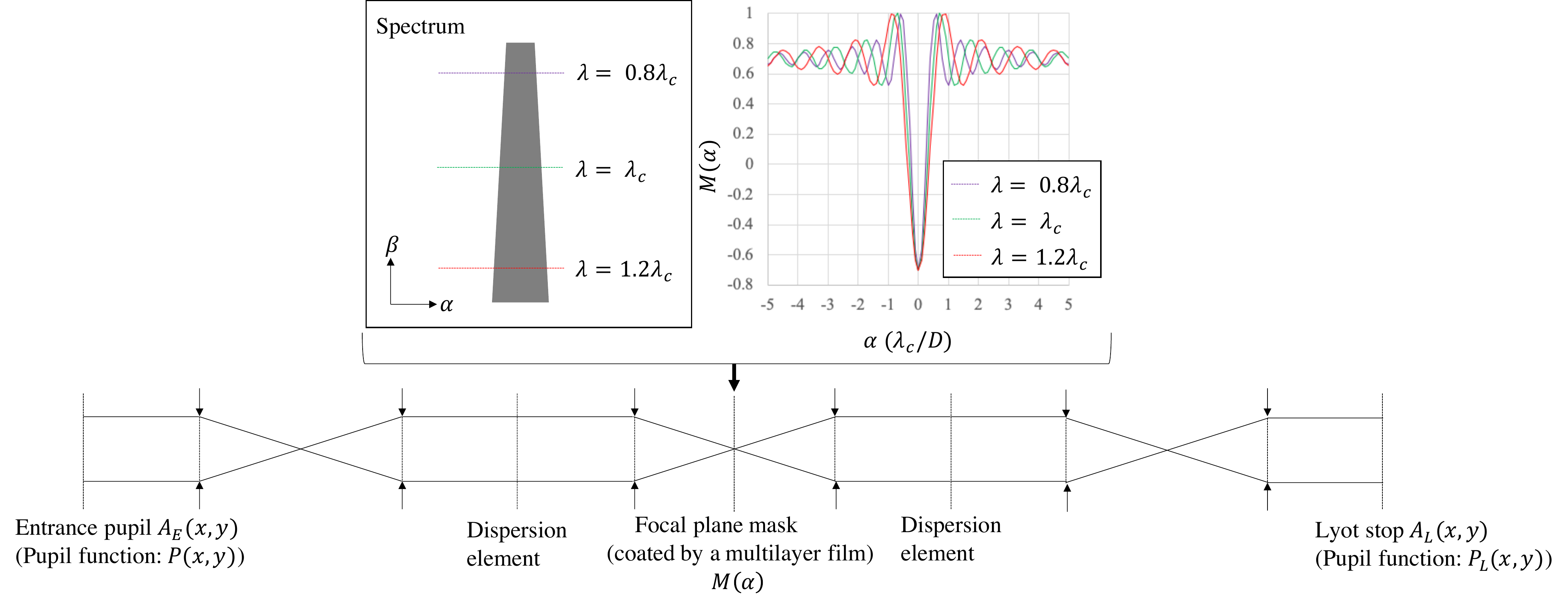}
	\caption{Conceptual diagram of the first-stage spectroscopic coronagraph. The vertical arrows represent collimators or camera mirrors. The left and right upper panels show a cartoon of the spectrum and the complex amplitudes of the focal-plane mask at three wavelengths, 0.8 $\lambda_{c}$, $\lambda_{c}$, and 1.2 $\lambda_{c}$, respectively. $w_{0}$ is 2, and there is no pupil obscuration.}
	\label{fig:concept_each_stage}
\end{figure}

\begin{figure}
	 \centering
	\includegraphics[scale=0.3,height=7cm,clip]{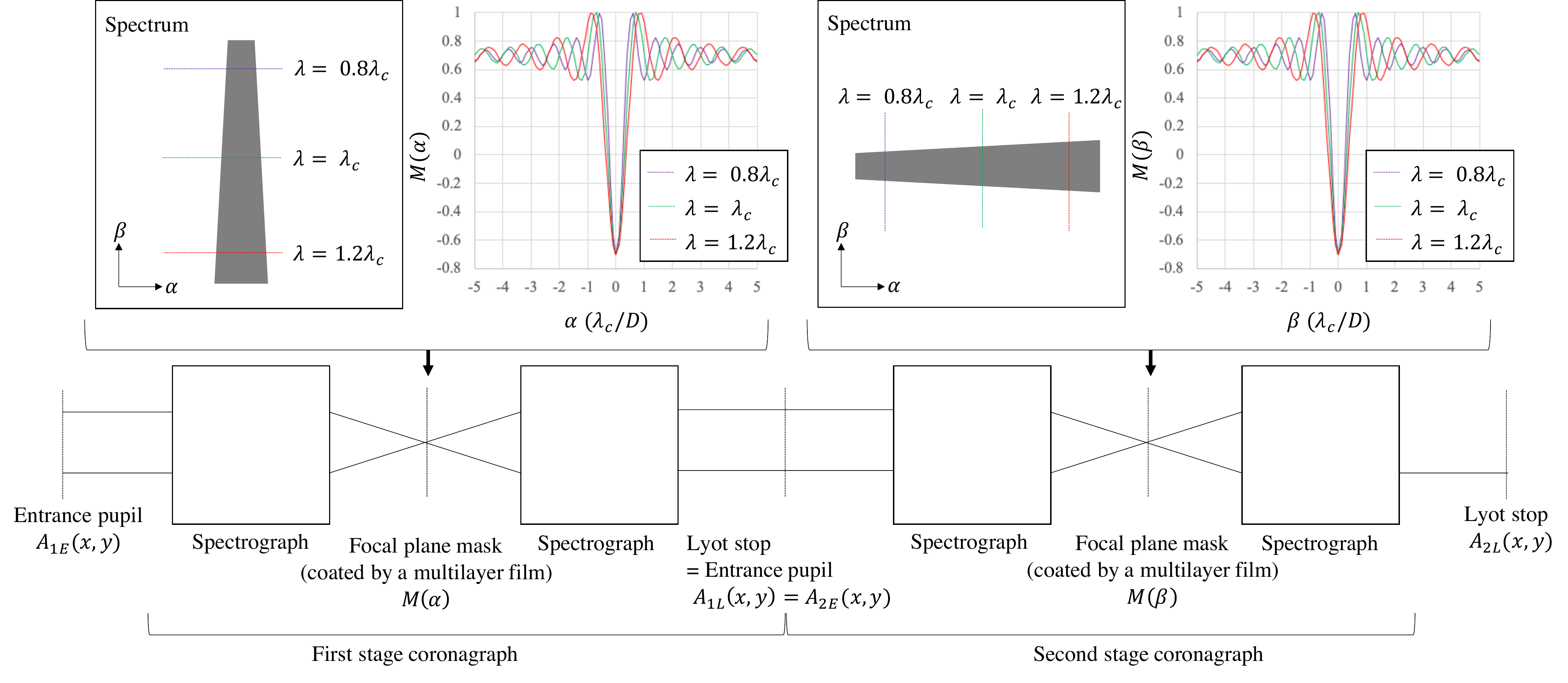}
	\caption{Block diagram of the spectroscopic fourth-order coronagraph. The complex amplitudes of the focal-plane mask at three wavelengths are the same as those shown in Figure \ref{fig:concept_each_stage}. Two second-order coronagraphs are placed in sucession, orthogonal to each other.}	
	\label{fig:concept}
\end{figure}

Here, focusing on the fact that this coronagraph mask modulates the complex amplitude along one direction on the focal plane, we noticed that it is possible to null the on-axis source over a wide bandwidth by applying a spectrograph to this coronagraph and realizing the modulation function optimized for the spectrally-resolved Airy disk. We name this coronagraph system, "spectroscopic fourth-order coronagraph." Figure \ref{fig:concept_each_stage} shows the conceptual diagram of the spectroscopic fourth-order coronagraph. Moreover, placing two spectroscopic coronagraphs in succession, parallel or orthogonal to each other, affords a fourth-order coronagraph, as shown in Figure \ref{fig:concept}. Note that the previous coronagraphs applying the complex amplitude masks on the focal plane \citep{Guyon+2010, Guyon+2014} achieve the second-order null and will be more affected by the telescope pointing jitter and finite stellar angular diameter for large telescopes. In this paper, we assume that the first- (second-) stage spectroscopic coronagraph generates the spectrum along the $\beta (\alpha)$ axis and modulates the complex amplitude along the $\alpha (\beta)$ axis. The first exit pupil, on which the Lyot stop is placed, corresponds to the entrance pupil of the latter coronagraph. We note that the coronagraph performance is not largely degraded even if the two coronagraphs are placed paralell to each other.

Conversely, new problems arise with using this spectroscopic coronagraph. Since the wavelength of each spectral element determines the modulation function of the complex amplitude, the modulation function weakly depends on the spectral direction (i.e., the $\beta$ direction for the first-stage coronagraph). In other words, the modulation function cannot be described by a variable of one axis; $m(\alpha_{\frac{\lambda}{D}})$ should change to $m(\alpha_{\frac{\lambda}{D}}, \beta_{\frac{\lambda}{D}})$. Consequently, the stellar leak is slightly generated by the optimized mask for the spectroscopic coronagraph. Furthermore, because the optical path is not common over the observation bandwidth, chromatic aberration, due to the non-common path errors that cannot be compensated by deformable mirrors in the upstream section of the coronagraph system, degrades the contrast. Based on these considerations, we investigate the impact of the optimized modulation function on the performance of the coronagraph in Section \ref{subsec:mask}, and in Section \ref{subsec:propagation}, we analytically describe the propagation of the non-common path errors through the coronagraph system with a fourth-order null.  

\subsection{Focal-plane mask} \label{subsec:mask}
The modulation period of the focal-plane mask applied to this spectroscopic coronagraph should be optimized for the spectrally-resolved wavelength; $\frac{\lambda}{\lambda_{c}}$ of Equation \ref{tilde_mask} becomes 1. Consequently, the modulation period should continuously change along the direction of the spectrum formed on the focal plane. For the first-stage coronagraph, the complex amplitude is modulated along the $\alpha$ axis, and the spectrum is formed along its perpendicular axis (i.e., $\beta$ axis) (see Section \ref{subsec:overview}). Considering that the Airy disk's diameter changes along the $\beta$ axis, the optimum modulation period should be proportional along the same axis. In this case, the modulation function of the focal-plane mask in the first-stage coronagraph, $m_{spectrum,1}(\alpha_{\frac{\lambda}{D}},\beta_{\frac{\lambda}{D}})$, is expressed as 
\begin{equation}
	\label{equ:m_spec}
	m_{spectrum,1}(\alpha_{\frac{\lambda}{D}},\beta_{\frac{\lambda}{D}}) = \frac{w_{0}}{\xi_{x}} \mathrm{sinc}\left \{w_{0}\pi \left(\alpha_{\frac{\lambda}{D}} + \chi \beta_{\frac{\lambda}{D}}\right)\right \},
\end{equation}
where $\chi$ is the change amount of the modulation period when $\beta$ shifts to $\beta + \delta \beta$. Given that the difference between the focus points at $\lambda$ and $\lambda + \delta \lambda$ is equal to half of the Airy disk's diameter, $\frac{\lambda}{D}$, $\chi$ is given as
\begin{equation}
	\chi = \frac{\delta \lambda}{\lambda} \equiv \frac{1}{R},
\end{equation}
where $R$ represents the resolving power of the spectrograph. Note that the $R$ value applied by this study is the same as the conventional definition of spectral resolution; adjacent spectral elements can be resolved when the distance between the elements is more than 1 $\frac{\lambda}{D}$. When the resolving power is more than a few hundred, the $\chi\beta_{\frac{\lambda}{D}}$ term is considered to be small unless $\beta_{\frac{\lambda}{D}}$ is large. Equation \ref{equ:m_spec} is expanded as follows: 
\begin{eqnarray}
	\label{m_spe_approx}
	m_{spectrum,1}(\alpha_{\frac{\lambda}{D}},\beta_{\frac{\lambda}{D}}) &\simeq & m_{1}(\alpha_{\frac{\lambda}{D}}) + \chi \beta_{\frac{\lambda}{D}} \left. \frac{d m_{1}(\alpha_{\frac{\lambda}{D}})}{d \alpha_{\frac{\lambda}{D}}} \right|_{\chi \beta_{\frac{\lambda}{D}} = 0} \nonumber \\
	&=&  m_{1}(\alpha_{\frac{\lambda}{D}}) + \frac{w_{0}}{\xi_{x}} \left( \frac{w_{0}\pi \beta_{\frac{\lambda}{D}}}{R} \right)\left\{ \frac{\cos\left(w_{0}\pi \alpha_{\frac{\lambda}{D}}\right)}{w_{0}\pi \alpha_{\frac{\lambda}{D}}} - \frac{\mathrm{sinc}\left(w_{0}\pi \alpha_{\frac{\lambda}{D}}\right)}{w_{0}\pi \alpha_{\frac{\lambda}{D}}} \right\}.
\end{eqnarray}
The second term of the right-hand side of Equation \ref{m_spe_approx} represents the difference between the ideal modulation function and the one applied to the spectroscopic coronagraph. We define the difference as $\Delta m_{1}(\alpha_{\frac{\lambda}{D}}, \beta_{\frac{\lambda}{D}})$. Note that the approximation shown in Equation \ref{m_spe_approx} should be expanded by higher order Taylor series if $\chi \beta_{\frac{\lambda}{D}}$ is not considered to be fully smaller than $\alpha_{\frac{\lambda}{D}}$, and the second-order Taylor series expansion of the modulation function is shown in Appendix \ref{sec:appendix_a}. Since the estimation of the stellar leak is not affected by the higher order expansion series under a certain condition (see Appendix \ref{sec:appendix_a}), the first-series Taylor expansion is used. 

\begin{figure}
	 \centering
	\includegraphics[scale=0.3,height=7cm,clip]{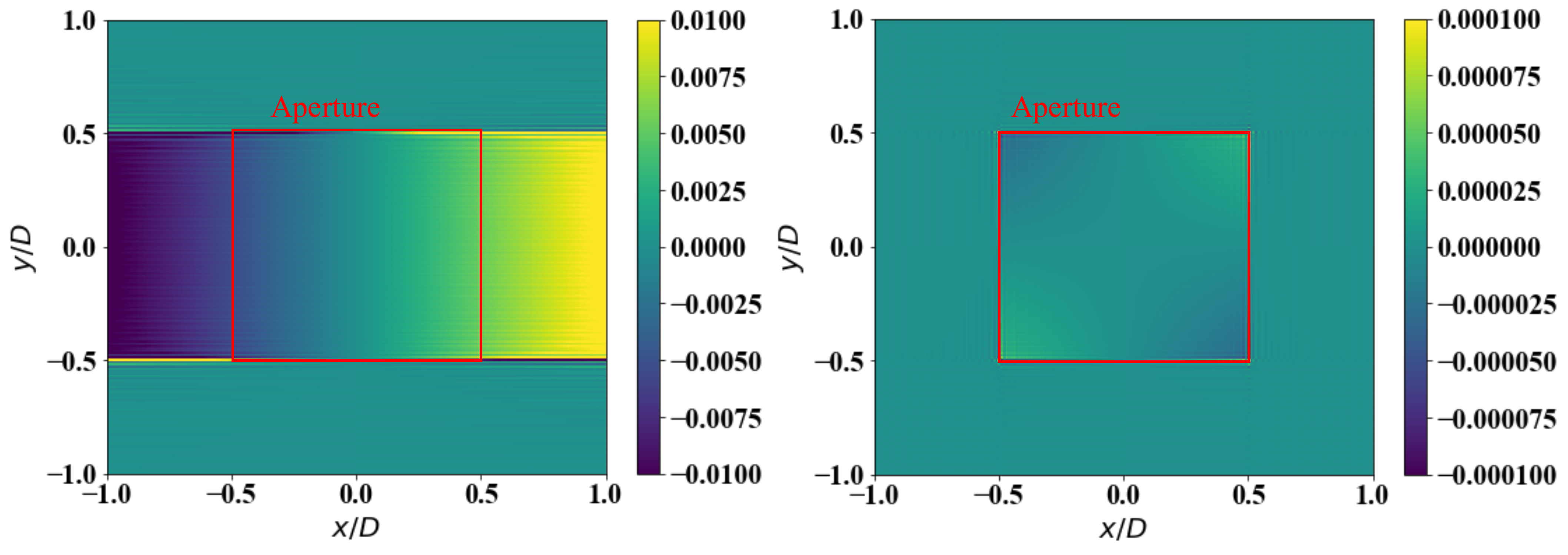}
	\caption{Residual amplitude on the Lyot stop for the light passing at the first- (left) and second-stage coronagraph (right). The spectral resolution, and length of the mask are set to 670 and 120 $\frac{\lambda}{D}$, respectively. Additionally, no obscuration of the aperture is assumed to exist, and the light passes through the center of the focal-plane mask. For both cases, the amplitude modulation functions of the mask for the first- and second-stage coronagraphs, $m_{spectrum,1}(\alpha_{\frac{\lambda}{D}},\beta_{\frac{\lambda}{D}})$ and $m_{spectrum,2}(\alpha_{\frac{\lambda}{D}},\beta_{\frac{\lambda}{D}})$, satisfy Equations \ref{m_spe_approx} and \ref{m_spectrum_2}, respectively.}
	\label{fig:A_residual_lyot_stop}
\end{figure}

The convolution of the Fourier conjugate of $\Delta m_{1}(\alpha_{\frac{\lambda}{D}}, \beta_{\frac{\lambda}{D}})$ with the entrance pupil generates the unwanted stellar leak. Since $\Delta m_{1}(\alpha_{\frac{\lambda}{D}}, \beta_{\frac{\lambda}{D}})$ is relatively large for the wide wavelength range, it is better to restrict the observation bandwidth (i.e., the length of the mask) to reduce the stellar leak. When the origin of the coordinate system on the focal plane, $(\alpha_{\frac{\lambda}{D}}, \beta_{\frac{\lambda}{D}})$, is the focal point of each spectrally-resolved light, the mask ranges from $-B_{-\frac{\lambda}{D}}$ to $B_{+\frac{\lambda}{D}}$ along the spectral direction. Note that the origin of the coordinate system changes with the wavelength. The Fourier conjugate of $\Delta m_{1}(\alpha_{\frac{\lambda}{D}}, \beta_{\frac{\lambda}{D}})$ is 
\begin{eqnarray}
	\label{delta_m}
	\Delta \tilde{m}_{1}(x,y,\lambda) &=& \int_{-\infty}^{\infty} d\alpha_{\frac{\lambda}{D}} \int_{-B_{-\frac{\lambda}{D}}}^{B_{+\frac{\lambda}{D}}}  d\beta_{\frac{\lambda}{D}} \Delta m_{1}(\alpha_{\frac{\lambda}{D}}, \beta_{\frac{\lambda}{D}}) \mathrm{e}^{ - 2\pi i \left(\alpha_{\frac{\lambda}{D}} \frac{x}{D}  + \beta_{\frac{\lambda}{D}} \frac{y}{D} \right)}\nonumber \\
	&=& \frac{2 i}{R \xi_{x}} \left(\frac{x}{D} \right) \mathrm{rect}\left( \frac{x}{w_{0} D}\right) \left \{ -\frac{B_{+\frac{\lambda}{D}} \mathrm{e}^{-2 \pi i B_{+\frac{\lambda}{D}} \frac{y}{D}} + B_{-\frac{\lambda}{D}} \mathrm{e}^{2\pi i B_{-\frac{\lambda}{D}} \frac{y}{D}}}{{\left(2 \pi i \frac{y}{D} \right)}} + \frac{\mathrm{e}^{-2 \pi i B_{+\frac{\lambda}{D}} \frac{y}{D}} - \mathrm{e}^{2 \pi i B_{-\frac{\lambda}{D}} \frac{y}{D}}}{{\left(2 \pi \frac{y}{D} \right)^{2}}}\right \}.
\end{eqnarray}
Here, given that the light of the central wavelength passes through the center of the mask, $B_{-\frac{\lambda}{D}}$ is equal to $B_{+\frac{\lambda}{D}}$ for that wavelength, and only the real part remains in $\Delta \tilde{m}_{1}(x,y,\lambda)$ shown in Equation \ref{delta_m}. When the length of the mask is set to $2B_{c}\frac{\lambda_{c}}{D}$, Equation \ref{delta_m} for the central wavelength is rewritten as 
\begin{equation}
	\label{delta_m1}
	\Delta \tilde{m}_{1}(x,y,\lambda = \lambda_{c}) = \frac{4 B_{\frac{\lambda_{c}}{D}} }{R \xi_{x}} \left(\frac{x}{D} \right) \mathrm{rect}\left( \frac{x}{w_{0} D}\right) \left \{ \frac{\mathrm{sinc} \left( 2 \pi B_{\frac{\lambda_{c}}{D}} \frac{y}{D} \right) - \cos \left( 2 \pi B_{\frac{\lambda_{c}}{D}} \frac{y}{D} \right)}{2\pi \frac{y}{D}} \right \}. 
\end{equation}
The residual complex amplitude propagating through the Lyot stop at the central wavelength, $A_{1L,mask}(x,y,\lambda = \lambda_{c})$, is derived from the convolution of $\Delta \tilde{m}_{1}(x,y,\lambda = \lambda_{c})$ with the entrance pupil, $P(x)P(y)$, along the $x$ direction. Given that $w_{0}$ is equal to 2 and the Lyot stop is the same as the entrance pupil, the residual complex amplitude on the first Lyot stop, $A_{1L,mask}(x,y,\lambda = \lambda_{c})$, is
\begin{eqnarray}
	\label{A_res_pupil}
	A_{1L,mask}(x,y,\lambda = \lambda_{c}) &=& P_{1}(x)P_{1}(y) \left\{\Delta \tilde{m}_{1}(x,y,\lambda=\lambda_{c}) * P_{1}(x)P_{1}(y) \right\} \nonumber \\
		           &=& \frac{4 B_{\frac{\lambda_{c}}{D}}}{R} P_{1}(x) \left \{ \frac{1}{\xi_{x}} \left(\frac{x}{D} \right) \mathrm{rect}\left( \frac{x}{2D}\right)  * \left((1-P_{s,x}(x)) \mathrm{rect}\left(\frac{x}{D}\right) \right)  \right \}	\nonumber \\
		           &\quad & \times P_{1}(y) \left \{ \left( \frac{\mathrm{sinc} \left( 2 \pi B_{\frac{\lambda_{c}}{D}} \frac{y}{D} \right) - \cos \left( 2 \pi B_{\frac{\lambda_{c}}{D}} \frac{y}{D} \right)}{2 \pi \left(\frac{y}{D} \right)} \right) * \left( (1-P_{s,y}(y)) \mathrm{rect}\left(\frac{y}{D}\right) \right) \right \} \nonumber \\		           
		           &=& \frac{2 B_{\frac{\lambda_{c}}{D}}}{R} \left(\frac{x}{D} \right) P_{1}(x)P_{1}(y) \left \{ \left( \frac{\mathrm{sinc} \left( 2 \pi B_{\frac{\lambda_{c}}{D}} \frac{y}{D} \right)  - \cos \left( 2 \pi B_{\frac{\lambda_{c}}{D}} \frac{y}{D} \right)}{2\pi \left( \frac{y}{D}\right)} \right)  * P_{1}(y) \right \} \nonumber \\
		           &=& U_{y}\left(B_{\frac{\lambda_{c}}{D}}, R\right) \left(\frac{x}{D} \right) P_{1}(x)P_{1}(y),               
\end{eqnarray}
where $U_{i}$ represents the constant value originating from the convolution of $\Delta \tilde{m}(x,y,\lambda=\lambda_{c})$ with the pupil function along the spectral direction ($i$-axis) and is determined by $B_{\frac{\lambda_{c}}{D}}$ and $R$; the $i$-axis corresponds to the $y$- and $x$-axes in the first- and second-stage coronagraph, respectively. The reason why $U_{i}$ is constant is that $B_{\frac{\lambda_{c}}{D}}\frac{\mathrm{sinc} \left( 2 \pi B_{\frac{\lambda_{c}}{D}} \frac{y}{D} \right)  - \cos \left( 2 \pi B_{\frac{\lambda_{c}}{D}} \frac{y}{D} \right)}{2 \pi \left(\frac{y}{D}\right)} $ rapidly increases at approximately $y=0$ and approaches $B_{\frac{\lambda_{c}}{D}}^{2}$ at $y=0$. The left image of Figure \ref{fig:A_residual_lyot_stop} shows the residual amplitude on the Lyot stop under the assumption that the resolving power of the spectrograph, $R$, and the half-length of the mask, ${B_{\frac{\lambda_{c}}{D}}}$, are set to 670 and 60, respectively. $U_{y}(B_{\frac{\lambda_{c}}{D}}=60, R=670)$ is approximately 0.01 under this condition.

Next, we investigate how the residual amplitude on the Lyot stop of the first-stage coronagraph passes through the second-stage coronagraph. Given that the second-stage coronagraph has the same optical parameters as that of the first-stage one, the modulation function of the focal-plane mask in the second-stage coronagraph is expressed as
\begin{eqnarray}
	\label{m_spectrum_2}
	m_{spectrum,2}(\alpha_{\frac{\lambda}{D}},\beta_{\frac{\lambda}{D}}) &=& \frac{w_{0}}{\xi_{x}} \mathrm{sinc}\left \{w_{0}\pi \left(\beta_{\frac{\lambda}{D}} + \chi \alpha_{\frac{\lambda}{D}}\right)\right \} \nonumber \\
	&\simeq & m_{2}(\beta_{\frac{\lambda}{D}}) + \Delta m_{2}(\alpha_{\frac{\lambda}{D}}\beta_{\frac{\lambda}{D}}).
\end{eqnarray}
The residual amplitude on the second Lyot stop is described as the convolution of the modulation function of the focal-plane mask with the residual amplitude on the first one: 
\begin{eqnarray}
	\label{A_res_pupil_2}
	A_{2L,mask}(x,y,\lambda = \lambda_{c}) &=& P_{1}(x)P_{1}(y) (A_{1L,mask}(x,y,\lambda = \lambda_{c}) \nonumber \\
	&\quad & - \left(\tilde{m}_{2}(x,y,\lambda=\lambda_{c})+ \Delta \tilde{m}_{2}(x,y,\lambda=\lambda_{c}) \right) * A_{1L,mask}(x,y,\lambda = \lambda_{c}) ) \nonumber \\
\end{eqnarray}
where the second Lyot stop is assumed to be the same as the first one. Focusing on the fact that the residual amplitude on the first Lyot stop is constant along the modulation direction of the focal-plane mask in the second-stage coronagraph, the convolution of the Fourier conjugate of the modulation function with the residual amplitude on the first Lyot stop becomes 0: $\tilde{m}_{2}(x,y,\lambda=\lambda_{c})*A_{1L,mask}(x,y,\lambda=\lambda_{c}) = 0$. Therefore, Equation \ref{A_res_pupil_2} is rewritten as 
\begin{eqnarray}
	\label{A_res_pupil_2_result}
	A_{2L,mask}(x,y,\lambda = \lambda_{c}) &=& P_{1}(x)P_{1}(y) (\Delta \tilde{m}_{2}(x,y,\lambda=\lambda_{c}) * A_{1L,mask}(x,y,\lambda = \lambda_{c}) ) \nonumber \\
	&=& U_{x}\left(B_{\frac{\lambda_{c}}{D}}, R\right) U_{y}\left(B_{\frac{\lambda_{c}}{D}}, R\right) \left(\frac{x}{D} \right) \left(\frac{y}{D} \right) P_{1}(x)P_{1}(y) ,
\end{eqnarray}
The right image of Figure \ref{fig:A_residual_lyot_stop} shows the residual amplitude on the Lyot stop of the second-stage coronagraph. The two coronagraphic masks generate the cross-term of the two tilts along the $x$- and $y$-axes. For the other wavelengths, the imaginary part, as well as the real part, remains in Equation \ref{delta_m1} because the light does not pass through the coronagraphic mask center. $B_{\frac{\lambda_{c}}{D}}^{2}$ at $y=0$ for the center wavelength changes into $\frac{B_{+\frac{\lambda}{D}}^{2} + B_{-\frac{\lambda}{D}}^{2}}{2}$ and $\frac{B_{+\frac{\lambda}{D}}^{2} - B_{-\frac{\lambda}{D}}^{2}}{2}$ for the real and imaginary parts, respectively. Therefore, the stellar leak increases as the wavelength further deviates from the central wavelength, $\lambda_{c}$. 

Since the residual amplitude on the Lyot stop is proportional to the square of the length of the mask, $B_{\frac{\lambda}{D}}^{2}$, the length of the mask should be adjusted according to the target contrast of each instrument; the observation bandwidth is more limited as the target contrast is higher. However, the strong dependence of the length of the mask on the stellar leak should be carefully treated for a large focal-plane mask (i.e., large $B_{\frac{\lambda}{D}}$) because the coronagraphic mask is not analytically approximated well; the stellar leak may weakly depend on the length of the mask for the large $B_{\frac{\lambda}{D}}$. For the central wavelength, the complex amplitude of the stellar light passing through the edge of the focal-plane mask (i.e., 100 $\frac{\lambda}{D}$) is much smaller than its peak at the center of the mask. Therefore, the modulation function gives a negligible impact on the complex amplitude of the stellar light at the large $\frac{\lambda}{D}$. In other words, this analytical approximation of the modulation function provides the lower limit of the observation bandwidth. As the next step of this study, we need to perform numerical simulations to evaluate the observation bandwidth through investigating the dependence of the length of the mask on the stellar leak for the large $B_{\frac{\lambda}{D}}$. 

Although the dependence of the length of the mask on the stellar leak may be weak at the large $B_{\frac{\lambda}{D}}$, we present a solution for mitigating the impact of the modulation function on the coronagraph performance at that region, assuming that Equation \ref{A_res_pupil_2_result} is valid for the large $B_{\frac{\lambda}{D}}$. Focusing on the fact that a linear variable filter allows the light of each wavelength to pass through a different position on the filter, we noticed that the linear variable filter could restrict the optimum length of the mask along the spectral direction for each wavelength. Furthermore, the light of each wavelength passes through the center of the mask. Placing a linear variable filter on the focal plane suppresses the stellar leak due to the new coronagraphic mask and broadens the observation bandwidth. Conversely, when the linear variable filter independently is placed in front of the focal-plane mask, the gap between the filter and the focal-plane mask generates a ghost light. The spherical aberration is also formed due to the linear variable filter in the converging light. Therefore, we assume that a multilayer film working as the linear variable filter is applied to the substrate of the focal-plane mask. Note that a linear variable filter could be manufactured as a layer of the focal-plane array \citep[e.g.,][]{Ahlberg+2017}, and various types of coatings have been used for the focal-plane masks so far \citep[e.g.,][]{Mawet+2009,Galicher+2020}. 

In Section \ref{sec:spectroscopic_coronagraph_design}, we will evaluate the stellar leak \replaced{occurred}{occurring} due to the new coronagraphic masks for an optimized optical design of the spectroscopic coronagraph using Equation \ref{A_res_pupil_2_result}.

\subsection{Propagation of Chromatic Aberration} \label{subsec:propagation}
We investigate the effect of the high-order chromatic aberrations (i.e., non-common path error) generated in the spectroscopic fourth-order coronagraph on the stellar leak, analytically describing their propagation through the coronagraph system with a fourth-order null. The high-order chromatic aberration is formed in the optical path between two dispersion elements before/after the coronagraph mask, as shown in Figure \ref{fig:concept_each_stage}. The aberrations are mainly divided into two in terms of their impact on the coronagraph performance: (1) those generated between the dispersion element and the camera system before the coronagraphic mask and (2) those generated between the collimator system and the dispersion element after the coronagraphic mask. The former and latter are formed close to the entrance and exit pupils. Since the latter aberration multiplies the entire right side of Equation \ref{amp_lyot}, the stellar leak is not influenced by the latter but by the former. Therefore, the chromatic aberration formed between the dispersion element and the camera lens before the coronagraph mask is considered hereinafter. 

The high-order chromatic aberration formed in the $k$-th coronagraph system, $\phi_{k}(x,y,\lambda)$, is written, using $x$- and $y$-dependent functions and their cross-term function:  
\begin{equation}
	\label{phi}
	\phi_{k}(x,y,\lambda) = \phi_{k,x}(x,\lambda) + \phi_{k,y}(y,\lambda) + \phi_{k,xy}(x,\lambda)\phi_{k,xy}(y,\lambda)
\end{equation}
The complex amplitude on the entrance pupil of the first-stage coronagraph, $A_{1E}(x,y)$, is given as
\begin{eqnarray}
	\label{complex_amplitude_1}
	A_{1E,ncp}(x,y,\lambda) &=& P_{1}(x,y) (1 + i\phi_{1}(x,y,\lambda)) \nonumber \\
	&\simeq & P_{1}(x)P_{1}(y)\{1 + i(\phi_{1,x}(x,\lambda)+\phi_{1,y}(y,\lambda) + \phi_{1,xy}(x,\lambda)\phi_{1,xy}(y,\lambda))\},
\end{eqnarray}
where the second or higher-order aberrations are ignored because the wavefront aberration is expected to be significantly smaller than the first-order ones. The complex amplitude on the exit pupil of the first-stage coronagraph system having a chromatic aberration of $\phi_{1}(x,y,\lambda)$ is
\begin{eqnarray}
	\label{complex_amplitude_1L}
	A_{1L,ncp}(x,y,\lambda) &=& P_{L}(x,y)(A_{1E}(x,y,\lambda) - \tilde{m}(x,\lambda) * A_{1E}(x,y,\lambda)) \nonumber \\
	 &=& P_{1}(x,y)(P_{1}(x) - \tilde{m}(x,\lambda) * P_{1}(x)) \nonumber \\
	 &\quad & + iP_{1}(x,y)(P_{1}(x)\phi_{1,x}(x) - \tilde{m}(x,\lambda) * P_{1}(x)\phi_{1,x}(x))  \nonumber \\
	 &\quad & + iP_{1}(x,y)(P_{1}(y)\phi_{1,y}(y) - \tilde{m}(x,\lambda) * P_{1}(y)\phi_{1,y}(y))  \nonumber \\ 
	 &\quad & + iP_{1}(x,y)\phi_{1,xy}(y)(P_{1}(x)\phi_{1,xy}(x) - \tilde{m}(x,\lambda) * P_{1}(x)\phi_{1,xy}(x)), 
\end{eqnarray}
where the Lyot stop on the exit pupil, $P_{L}(x,y)$, is assumed to be identical to that on the input pupil, $P_{1}(x,y)$. Given that the coronagraphic mask could be optimized for the spectrally-resolved light, no stellar leak due to the non-high-order chromatic aberration (i.e., the first-term of the right-hand in Equation \ref{complex_amplitude_1L}) exists on the Lyot plane: $P_{1}(x) - \tilde{m}(x,\lambda) * P_{1}(x) = 0$. In addition, considering that $\tilde{m}(x,\lambda) * \phi_{1,y}(y) =  \phi_{1,y}(y)$, $\phi_{1,y}(y) - \tilde{m}(x,\lambda) * \phi_{1,y}(y) = 0$. In other words, the aberration function of the variable perpendicular to the modulation direction of the coronagraphic mask does not generate the stellar leak. Therefore, Equation \ref{complex_amplitude_1L} is rewritten as 
\begin{equation}
	A_{1L,ncp}(x,y,\lambda) = iP_{1}(x,y)\left\{(P_{1}(x)\phi_{1,x}(x) - C_{x}) +\phi_{1,xy}(y)(P_{1}(x)\phi_{1,xy}(x) - C_{xy}) \right\},
\end{equation}
where $C_{i}$ is the convolution of the Fourier conjugate of the coronagraphic mask, $\tilde{m}(x,\lambda)$, with the aberration functions of variable $i$, $P_{1}(x)\phi_{1,i}(i)$, which becomes a constant value. The complex amplitude on the entrance pupil of the second-stage coronagraph, $A_{2E,ncp}(x,y,\lambda)$, is expressed as the multiplication of that on the exit pupil of the first-stage one with the chromatic aberration generated by the second-stage one: 
\begin{eqnarray}
	\label{complex_amplitude_2L}
	A_{2E,ncp}(x,y,\lambda) &=& iP_{1}(x,y) \left \{(\phi_{1,x}(x) - C_{x}) + \phi_{1,xy}(y)(\phi_{1,xy}(x) - C_{xy}) \right \}  \nonumber \\
	 &\quad & \times \{1 + i (\phi_{2,x}(x,\lambda) + \phi_{2,y}(y,\lambda) + \phi_{2,xy}(x,\lambda) \phi_{2,xy}(y,\lambda)) \} \nonumber \\
	 &\simeq &  iP_{1}(x,y) \left\{  (\phi_{1,x}(x) - C_{x}) + \phi_{1,xy}(y)(\phi_{1,xy}(x) - C_{xy}) \right\}, 
\end{eqnarray}
where the second-order terms are ignored; consequently, the complex amplitude on the entrance pupil of the second-stage coronagraph is the same as that on the exit pupil of the first-stage coronagraph. The complex amplitude on the exit pupil of the second-stage coronagraph, $A_{2L,ncp}(x,y,\lambda)$, is expressed as
\begin{equation}
	\label{complex_amplitude_2L_result}
	A_{2L,ncp}(x,y,\lambda) = iP_{1}(x,y)(\phi_{1,xy}(x) - C_{xy})(\phi_{1,xy}(y) - C^{'}_{xy}) ,
\end{equation}
where $C^{'}_{xy}$ is the convolution of the Fourier conjugate of the latter coronagraph mask, $\tilde{m}(y)$, with the aberration function, $P_{1}(y)\phi_{1,xy}(y)$. The following relation: $\phi_{1,x} - \tilde{m}(y) * \phi_{1,x} = 0$, was used in the above calculation. Thus, the purely one-axis dependent aberration functions, $\phi(x)$ and $\phi(y)$, are removed by the fourth-order coronagraph. In contrast, the cross-term of the $x$- and $y$-dependent aberrations, $\phi_{xy}(x)\phi_{xy}(y)$, can be transmitted through the fourth-order coronagraph. The cross-term limits the contrast on the detector plane.   

If the cross-term, $\phi_{xy}(x)\phi_{xy}(y)$, is not generated in the spectroscopic coronagraph system, the second-order aberration function should be considered. Equation \ref{complex_amplitude_1} is rewritten as
\begin{equation}
	A_{1E,ncp}(x,y,\lambda) \simeq P_{1}(x,y) \left\{1 + i(\phi_{1,x}(x,\lambda)+\phi_{1,y}(y,\lambda)) - \frac{1}{2} \left(\phi_{1,x}(x)^{2} + \phi_{1,y}(y)^{2} + 2\phi_{1,x}(x)\phi_{1,y}(y) \right) \right\}
\end{equation}
Considering that the aberration function written with a purely one-axis variable does not propagate through the fourth-order coronagraph, the complex amplitude on the exit pupil of the second-stage coronagraph is given as 
\begin{equation}
	\label{iedal_2L}
	A_{2L,ncp}(x,y,\lambda) = - P_{1}(x,y)(\phi_{1,x}(x) - C_{x})(\phi_{1,y}(y) - C^{'}_{y}).
\end{equation}
The above equation appears similar to Equation \ref{complex_amplitude_2L_result}. However, the above equation shows the propagation of the second-order terms of the aberration function, and its complex amplitude is much smaller than that of Equation \ref{complex_amplitude_2L_result}. 

Based on these considerations, the impact of the high-order chromatic aberration on the stellar leak can be significantly reduced by preventing the formation of the cross-term, $\phi_{xy}(x)\phi_{xy}(y)$, in the spectroscopic coronagraph system. In this case, the stellar leak is determined by the second-order aberration function. We focus on applying an Offner-type imaging spectrograph to this spectroscopic coronagraph because this type of spectrograph does not generate non-axis aberrations under the condition that the optical system has no alignment error. In the next section, we propose a spectroscopic coronagraph design with an Offner-type imaging spectrograph. 

\section{Spectroscopic Fourth-order Coronagraph with an Offner Type Spectrograph} \label{sec:spectroscopic_coronagraph_design}
As discussed in Section \ref{sec:theory}, the stellar leak due to the coronagraphic mask optimized for the spectroscopic fourth-order coronagraph gives an impact on the performance of the coronagraph. The cross-term of the $x$- and $y$-dependent aberration functions, $\phi(x)\phi(y)$, propagates in the fourth-order coronagraph system and limits the contrast on the detector plane. A combination of the coronagraph applying one-dimensional modulation mask with an Offner-type imaging spectrograph, which does not generate a non-axis aberration, is considered here. In this section, we show an optical design for a spectroscopic coronagraph with an Offner-type imaging spectrograph, analytically derive the achievable contrast on the focal plane, and evaluate how much the observation bandwidth can be increased for the optical design.

\subsection{Design} \label{subsec:optical_design}
The imaging spectrograph with a concave reflection grating \citep[e.g.,][]{Lobb+94, Lobb+1997}, referred to as "Offner-type imaging spectrograph" in this paper, is a derivative of an Offner relay optical system composed of two concentric spherical mirrors  \citep{Offner+1975}; one of the two spherical mirrors is replaced with a concave diffraction grating. The rotational symmetry of the Offner-type imaging spectrograph does not generate a non-axis aberration, including an on-axis chromatic aberration \citep[e.g.,][]{Kim+2014}; no cross-term of the $x$- and $y$-dependent aberration functions, $\phi(x)\phi(y)$, exists. The pupil is formed on the concave grating if the former optical system has a telecentric design.

\begin{figure}
	 \centering
	\includegraphics[scale=0.6,height=4.5cm,clip]{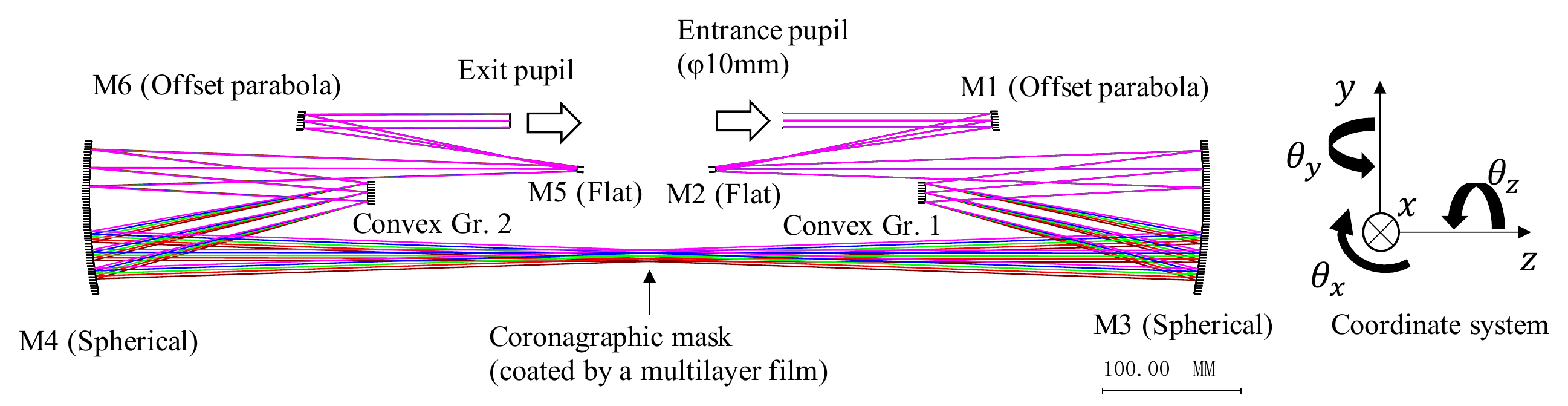}
	\caption{Optical design of the spectroscopic coronagraph with an Offner-type spectrograph designed based on the optical parameters shown in Table \ref{tab:parameters}. This system corresponds to the first-stage coronagraph system in Figure \ref{fig:concept}. While the spectrum image is produced along the $\beta$-axis parallel to $y$, the complex amplitude is modulated by the coronagraphic mask along the $\alpha$-axis. }
	\label{fig:optical_design}
\end{figure}

Figure \ref{fig:optical_design} shows the optical design of a spectroscopic coronagraph with an Offner-type imaging spectrograph. The optimized wavelength ranges from 600 to 800 $nm$, corresponding to a 30$\%$ bandwidth. A pupil mask suitable for the coronagraph design proposed by \cite{Itoh+2020} is placed on the entrance pupil; the pupil mask separates the two variables of the pupil function: $P(x,y)=P(x)P(y)$. The pupil mask is a square aperture with the size of 10 $mm$. The reflection grating is optically conjugated to the entrance pupil with a relay optical system composed of a parabolic mirror and a spherical mirror. The reflection grating with a groove density of 100 $lines/mm$ disperses the white light along the $y$ axis, corresponding to the spectral direction, and a spectrum is focused with an F-number of 15. Here, the Offner-type spectrograph was designed such that the defocus and astigmatism (0/90) do not generate at the central wavelength, $\lambda_{c}$, by adjusting the following two parameters: (1) the distance between the grating and spherical mirror and (2) the radius of the reflection grating. These aberrations are also minimized in the other wavelengths.

\begin{figure}
	 \centering
	\includegraphics[scale=0.6,height=6cm,clip]{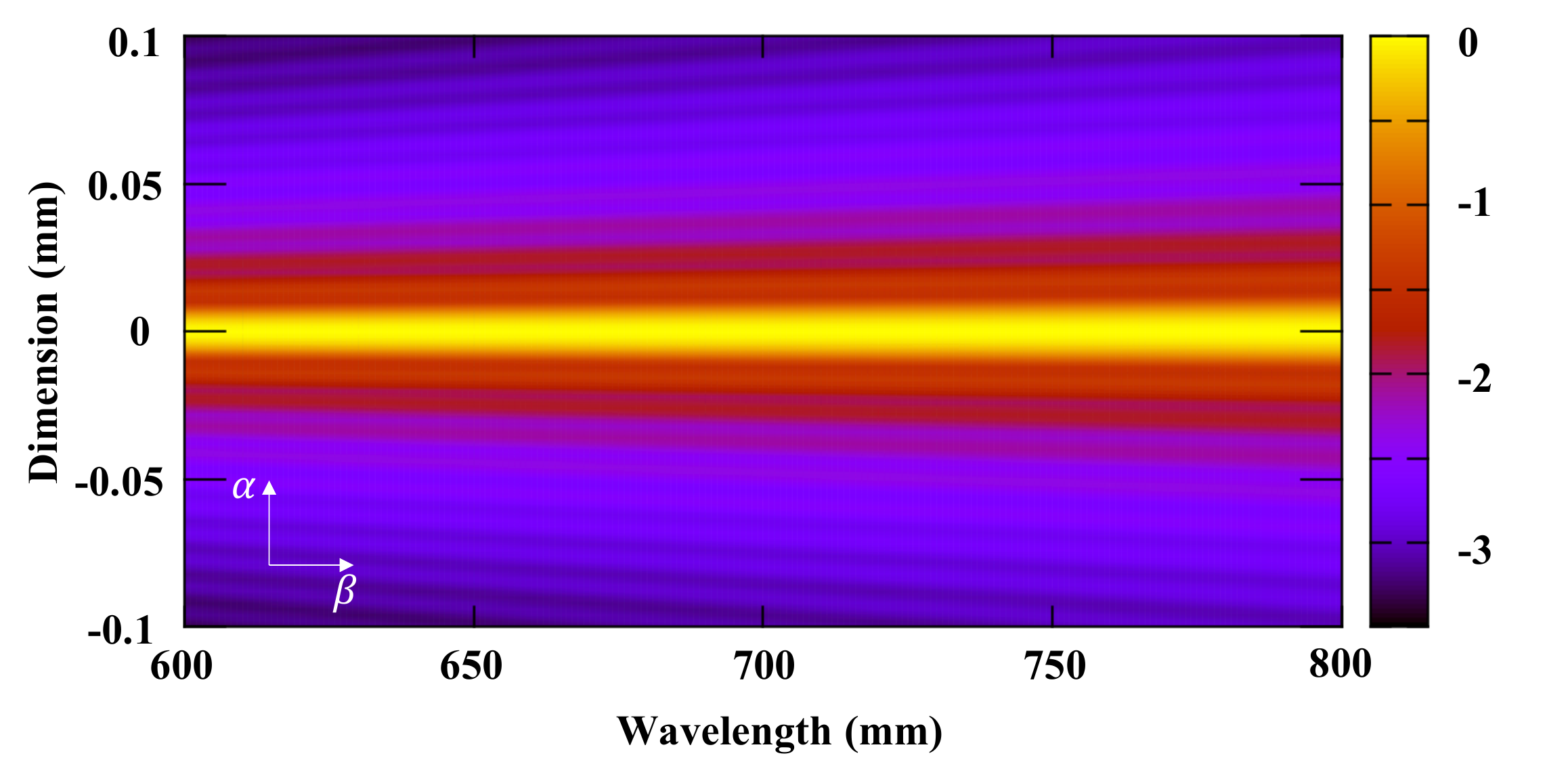}
	\caption{Focal-plane image produced by the Offner-type spectrograph shown in Figure \ref{fig:optical_design}. }
	\label{fig:focal_plane}
\end{figure}

As shown in Figure \ref{fig:focal_plane}, a spectrum with a spectral resolution of 667 is formed on the focal plane, on which the focal-plane mask is placed, and the spectral direction is along the $\beta$ axis parallel to the $y$-axis. The diameter of the formed Airy disk is 42 $\mu m$ at a central wavelength of 700 $nm$, and the coronagraphic mask modulates the complex amplitude within and outside the Airy disk. Note that one of the most promising methods for modulating the complex amplitude is to put a liquid-crystal-based phase waveplate between two linear polarizers; fortunately, the liquid-crystal-based phase waveplate has been developed for various applications, including the vector vortex coronagraph for long time \citep[e.g.,][]{Mawet+2011b, Tabirian+2015}. Another Offner-type imaging spectrograph, which has the same optical parameters as those used before the focal-plane mask, forms a white pupil on another reflection grating. Another white pupil is formed on a Lyot stop with another relay optics. The size of the entire coronagraph system is 850 x 150 x 20 (L x W x H) $mm$. Table \ref{tab:parameters} compiles the optical parameters of this system. We derive the contrast limited by the optimized focal-plane mask for the spectroscopic coronagraph and the high-order chromatic aberration in the following subsections.

\begin{table}[htb]
	\begin{center}
		\caption{Parameters of the spectroscopic coronagraph design}
  		\begin{tabular}{| l | c | r} \hline 
    		Item & Value  \\ \hline \hline
    		Type of spectrometer & Offner-type spectrometer \\ \hline
    		Wavelength coverage & 600 - 800 $nm$ \\ \hline
		Spectral resolution $\frac{\lambda}{\delta \lambda}$ & 667.3 \\ \hline
		Reduction ratio & 1.0 \\ \hline
		f ratio & 15 \\ \hline
		Radius of the spherical mirror & 400 $mm$ \\ \hline
		Radius of the convex grating & 200.303 $mm$ \\ \hline
     		Groove density of the convex grating & 50 lines/$mm$ \\ \hline
    		Distance between the mirror and the grating & 199.697 $mm$ \\ \hline
		Offset distance of the optical axis from the co-axis of the mirror and the grating & 17 $mm$ \\ \hline	
  \end{tabular}
  \label{tab:parameters}
  \end{center}
\end{table}

\subsection{Performance} \label{subsec:performance}
The performance of the spectroscopic coronagraph without any alignment errors of the optical elements is calculated. The impact of the alignment errors on the contrast performance will be introduced in Section \ref{sec:torelance}. This calculation utilizes a fourth-order coronagraph system composed of two second-order coronagraphs with the optical parameters shown in Table \ref{tab:parameters}. The incident wavefront to the spectroscopic coronagraph system is assumed to be perfect without any aberrations. In addition, the aberrations generated in the optical path between the entrance pupil and the reflection grating are not considered because the optical path is common over the wavelength range. We assume that the deformable mirrors in the upstream section of the coronagraph system correct the aberrations since they have a common optical path. However, as mentioned in Section \ref{sec:intro}, the periodic surface figure and reflectivity irregularities of nonpupil optics cannot be compensated by the deformable mirror on the pupil plane because the phase generated through the propagation of periodic wavefront errors has a linear wavelength dependency on the pupil plane \citep{Shaklan+2006}. Thus, the observation bandpass will be limited by the wavefront correction even if this spectroscopic coronagraph concept perfectly works in a wide bandpass.

Given that the optical elements are placed ideally, the stellar leak \replaced{occurred}{occurring} due to the new coronagraphic masks (Section \ref{subsec:mask}) and on-axial aberrations (Section \ref{subsec:propagation}), such as defocus and astigmatism (0/90), limit the contrast. The residual amplitude on the Lyot stop of the second-stage coronagraph is expressed as
\begin{equation}
	\label{A_second_lyot_stop_tot}
	A_{2L,tot}(x,y,\lambda) = A_{2L,mask}(x,y,\lambda) + A_{2L,ncp}(x,y,\lambda).
\end{equation}
The latter residual amplitude is originated from the following aberration function, $\phi(x,y)$:
\begin{equation}
	\label{chromatic_aberration}
	\phi(x,y) = \left (\frac{2\pi}{\lambda}\right)\left\{A \left\{\left(\frac{x}{D}\right)^{2} - \left(\frac{y}{D}\right)^{2}\right\} + 2B \left\{\left(\frac{x}{D}\right)^{2} + \left(\frac{y}{D}\right)^{2}\right\} \right\},
\end{equation}
where $A$ and $B$ are coefficients with the unit of the length. The first and second terms in the right-hand side of the above equation represent astigmatism (0/90) and defocus, respectively. The cross-term of the $x$- and $y$-dependent aberration functions, $\phi(x)\phi(y)$, does not exist, and the aberration function, $\phi(x,y)$, can be expressed with the $x$- and $y$-dependent functions, $\phi(x)$ and $\phi(y)$:
\begin{eqnarray}
	\label{phi_x_y}
	\phi(x)  &=& (2 \pi a_{\lambda}) \left( \frac{x}{D}\right)^{2} \\
	\phi(y) &=& (2 \pi b_{\lambda}) \left( \frac{y}{D} \right)^{2} , 
\end{eqnarray}
where $a_{\lambda}$ and $b_{\lambda}$ are defined as
\begin{eqnarray}
	a_{\lambda}  &=& \frac{A+2B}{\lambda} \\
	b_{\lambda} &=& \frac{2B-A}{\lambda} .
\end{eqnarray}

Next, we derived the coefficients, $A$ and $B$, from the wavefront map on the entrance pupil plane in the spectroscopic coronagraph design, as shown in Figure \ref{fig:optical_design}. The wavefront map was calculated through the Fourier transform of the complex amplitude on the focal plane. Table \ref{tab:aberration_ideal} shows the coefficients of five wavelengths, 600, 650, 700, 750, and 800 $nm$. The defocus and astigmatism (0/90) are perfectly removed at the central wavelength of 700 $nm$. Moreover, $a_{\lambda}$ becomes 0 at the two wavelengths, 600 and 800 $nm$, because the defocus and astigmatism (0/90) aberrations cancel it. 

\begin{table}[htb]
	\begin{center}
		\caption{High-order chromatic aberration without any alignment error}
  		\begin{tabular}{| l | c | c | c | c| c |} \hline 
    		Item & 600 $nm$ & 650 $nm$ & 700 $nm$ & 750 $nm$ & 800 $nm$\\ \hline \hline
    		Astigmatism (0/90) (coef. A) & 0.12  & 0.07 & 0.00 & 0.00 & -0.16 \\ \hline
    		Defocus (coef. B) & -0.06 & -0.07 & 0.00 & -0.08 & 0.08 \\ \hline
  \end{tabular}
  \label{tab:aberration_ideal}
  \end{center}
\end{table}

Based on the above evaluation, we derived the achievable contrast on the detector plane, on which faint planet light is detected. We write the complex amplitude on the detector plane (i.e., focal plane), $A_{D,tot}(\alpha_{\frac{\lambda}{D}},\beta_{\frac{\lambda}{D}},\lambda)$, by summing the two residual complex amplitudes due to the new mask and the high-order chromatic aberration on that plane, $A_{D,mask}(\alpha_{\frac{\lambda}{D}},\beta_{\frac{\lambda}{D}},\lambda)$ and $A_{D,ncp}(\alpha_{\frac{\lambda}{D}},\beta_{\frac{\lambda}{D}},\lambda)$:
\begin{equation}
	A_{D,tot}(\alpha_{\frac{\lambda}{D}},\beta_{\frac{\lambda}{D}},\lambda) = A_{D,mask}(\alpha_{\frac{\lambda}{D}},\beta_{\frac{\lambda}{D}},\lambda) + A_{D,ncp}(\alpha_{\frac{\lambda}{D}},\beta_{\frac{\lambda}{D}},\lambda). 
\end{equation}
Here, since the light of each wavelength passes through the center of the mask thanks to the linear variable filter applying to the focal-plane mask, as discussed in Section \ref{subsec:mask}, the residual amplitude due to the new coronagraphic mask for all of the wavelengths can be expressed by that at the central wavelength, $A_{D,mask}(\alpha_{\frac{\lambda}{D}},\beta_{\frac{\lambda}{D}},\lambda=\lambda_{c})$. Using Equations \ref{A_res_pupil_2_result} and \ref{iedal_2L}, $A_{D,mask}(\alpha_{\frac{\lambda}{D}},\beta_{\frac{\lambda}{D}},\lambda=\lambda_{c})$ and $A_{D,ncp}(\alpha_{\frac{\lambda}{D}},\beta_{\frac{\lambda}{D}},\lambda)$ are, respectively, expressed as
\begin{eqnarray}
	\label{A_focal_mask}
	A_{D,mask}(\alpha_{\frac{\lambda}{D}},\beta_{\frac{\lambda}{D}},\lambda=\lambda_{c}) &=& \int\int dxdy A_{2L,mask}(x,y,\lambda=\lambda_{c})  \mathrm{e}^{- 2\pi i \left(\frac{x}{D}\alpha_{\frac{\lambda}{D}}+\frac{y}{D}\beta_{\frac{\lambda}{D}} \right) } \nonumber \\ 
	&=& U_{x}\left(B_{\frac{\lambda_{c}}{D}}, R\right) U_{y}\left(B_{\frac{\lambda_{c}}{D}}, R\right) \int dx \left(\frac{x}{D} \right)(1-P_{s,x}(x))\mathrm{rect} \left(\frac{x}{D} \right) \mathrm{e}^{- 2\pi i \frac{x}{D}\alpha_{\frac{\lambda}{D}}}  \nonumber \\
	&\quad & \times \int dy \left(\frac{y}{D} \right)(1-P_{s,y}(y))\mathrm{rect}\left(\frac{y}{D} \right) \mathrm{e}^{- 2\pi i \frac{y}{D}\beta_{\frac{\lambda}{D}}} \nonumber \\
	&=& U_{x}\left(B_{\frac{\lambda_{c}}{D}}, R\right) U_{y}\left(B_{\frac{\lambda_{c}}{D}}, R\right) \left( \delta (\alpha_{\frac{\lambda}{D}}) - \tilde{P}_{s,x}(\alpha_{\frac{\lambda}{D}}) \right) * g_{\alpha, 1}(\alpha_{\frac{\lambda}{D}}) \nonumber \\
	&\quad & \times \left(  \delta (\beta_{\frac{\lambda}{D}}) - \tilde{P}_{s,y}(\beta_{\frac{\lambda}{D}}) \right) * g_{\beta, 1}(\beta_{\frac{\lambda}{D}}),
\end{eqnarray}	
and
\begin{eqnarray}
	\label{A_focal_ncp}
	A_{D,ncp}(\alpha_{\frac{\lambda}{D}},\beta_{\frac{\lambda}{D}},\lambda) &=& \int\int dxdy A_{2L,ncp}(x,y,\lambda=\lambda_{c})  \mathrm{e}^{- 2\pi i \left(\frac{x}{D}\alpha_{\frac{\lambda}{D}}+\frac{y}{D}\beta_{\frac{\lambda}{D}} \right) } \nonumber \\ 
	&=& i \int dx (1-P_{s,x}(x))\mathrm{rect} \left(\frac{x}{D} \right)(\phi_{1,x}(x) - C_{x}) \mathrm{e}^{- 2\pi i \frac{x}{D}\alpha_{\frac{\lambda}{D}}}  \nonumber \\
	&\quad & \times \int dy (1-P_{s,y}(y))\mathrm{rect}\left(\frac{y}{D} \right)(\phi_{1,y}(y) - C^{'}_{y})  \mathrm{e}^{- 2\pi i \frac{y}{D}\beta_{\frac{\lambda}{D}}} \nonumber \\
	&=& i \left( \delta (\alpha_{\frac{\lambda}{D}}) - \tilde{P}_{s,x}(\alpha_{\frac{\lambda}{D}}) \right) * \left( g_{\alpha, 2}(\alpha_{\frac{\lambda}{D}}) - C_{x} f_{\alpha}(\alpha_{\frac{\lambda}{D}}) \right) \nonumber \\
	&\quad & \times \left(  \delta (\beta_{\frac{\lambda}{D}}) - \tilde{P}_{s,y}(\beta_{\frac{\lambda}{D}}) \right) * \left( g_{\beta, 2}(\beta_{\frac{\lambda}{D}}) - C^{'}_{y} f_{\beta}(\beta_{\frac{\lambda}{D}}) \right),
\end{eqnarray}	
where $\tilde{P}_{s,i}$ shows the Fourier conjugate of the pupil obscuration function along the $i$ axis, and $f_{\alpha}$ shows the complex amplitude along the $\alpha$ axis on the focal plane for a pupil without any obscurations and aberrations:
\begin{equation}
	f_{\alpha}(\alpha_{\frac{\lambda}{D}}) \equiv \int dx \mathrm{rect}\left(\frac{x}{D}\right) \mathrm{e}^{- 2\pi i \frac{x}{D}\alpha_{\frac{\lambda}{D}}}.
\end{equation}
$g_{\alpha, n}$ is the complex amplitude along the $\alpha$ axis on the focal plane for the same pupil with an aberration function of $(\frac{x}{D})^{n}$ and can be expressed with $f_{\alpha}(\alpha_{\frac{\lambda}{D}})$ as follows:
\begin{equation}
	\label{g_n}
	g_{\alpha,n}(\alpha_{\frac{\lambda}{D}}) = \frac{1}{(-2\pi i)^{n}} \frac{d^{n}}{d\alpha^{n}_{\frac{\lambda}{D}}} f_{\alpha}(\alpha_{\frac{\lambda}{D}}).
\end{equation}

\begin{figure}
	 \centering
	\includegraphics[scale=0.6,height=4.8cm,clip]{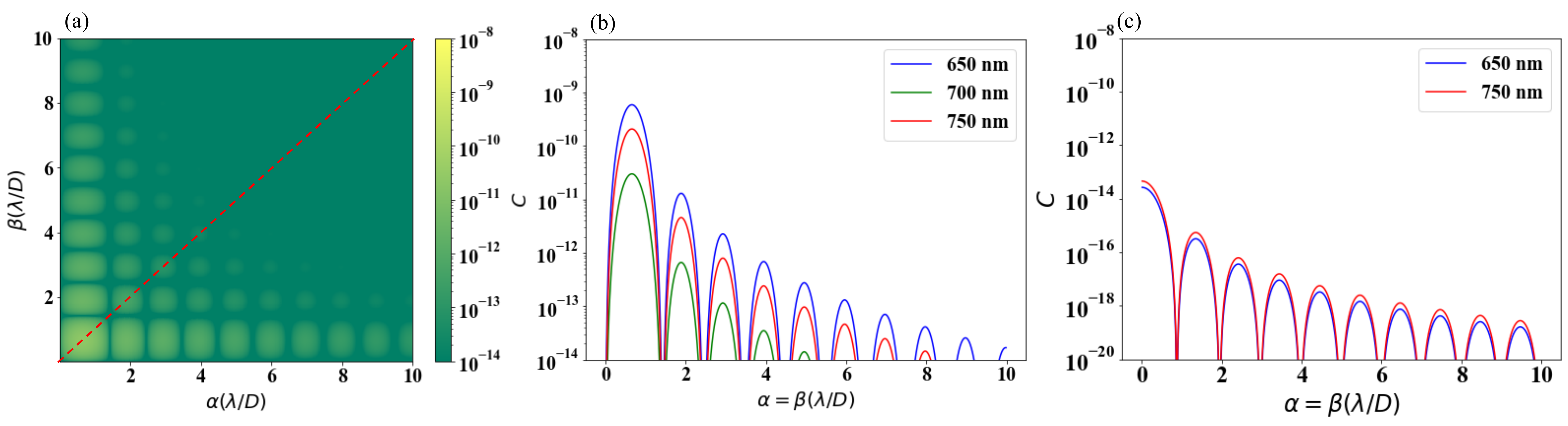}
	\caption{(a) Two-dimensional contrast map at the central wavelength of 700 $nm$ for the spectroscopic coronagraph with an Offner-type spectrograph applying the optical parameters shown in Table \ref{tab:parameters} (left). (b) Contrast curves limited by the new coronagraphic mask along $\alpha = \beta$ at three wavelengths, 650, 700, and 750 $nm$, $C_{D,mask}(\alpha_{\frac{\lambda}{D}}=\beta_{\frac{\lambda}{D}},\lambda=650,700,750~nm)$. The length of the mask is set to 120 $\frac{\lambda}{D}$ at 700 $nm$. While the light of the central wavelength passes through the center of the mask, the light at 650 and 750 $nm$ passes at offset distances of 46 and 52 $\frac{\lambda}{D}$ from the center of the mask. (c) Contrast curves limited by the high-order chromatic aberrations along $\alpha = \beta$ at two wavelengths, 650 and 750 $nm$, $C_{D,ncp}(\alpha_{\frac{\lambda}{D}}=\beta_{\frac{\lambda}{D}},\lambda=650, 750~nm)$. Given that the optical system has no alignment error, the high-order chromatic aberrations shown in Table \ref{tab:aberration_ideal} were applied to this contrast calculation.  The stellar leak does not occur at the central wavelength of 700 $nm$ because the coefficients of the aberration function, $a_{\lambda}$ and $b_{\lambda}$, are equal to 0.}
	\label{fig:ideal_limit}
\end{figure}

For the segmented telescope, the pupil obscuration function is determined by the gap between mirrors and is the same as the multiple narrow slits aligned at equal intervals. A number of the point-spread functions, $f_{\alpha}$ and $f_{\beta}$, are formed on the focal plane with the same effect as a grating \citep{Itoh+2019}. Since the pupil obscuration function does not affect the point-spread function close to the central star, we evaluate the spectroscopic coronagraph performance, assuming, hereinafter, that there is no obscuration on the pupil plane (i.e., $P_{s,x}(x) = P_{s,y}(y) = 0$ for $\forall x, y$). Based on the above considerations, Equations \ref{A_focal_mask} and \ref{A_focal_ncp} are rewritten as  
\begin{eqnarray}
	A_{D,mask}(\alpha_{\frac{\lambda}{D}},\beta_{\frac{\lambda}{D}},\lambda=\lambda_{c}) &=& U_{x}\left(B_{\frac{\lambda_{c}}{D}}, R\right) U_{y}\left(B_{\frac{\lambda_{c}}{D}}, R\right) g_{\alpha, 1}(\alpha_{\frac{\lambda}{D}})g_{\beta, 1}(\beta_{\frac{\lambda}{D}}) \nonumber \\
	&=& \frac{U_{x}\left(B_{\frac{\lambda_{c}}{D}}, R\right) U_{y}\left(B_{\frac{\lambda_{c}}{D}}, R\right)}{4\pi^{2}} \frac{d}{d\alpha_{\frac{\lambda}{D}}} f_{\alpha}(\alpha_{\frac{\lambda}{D}}) \frac{d}{d \beta_{\frac{\lambda}{D}}} f_{\beta}(\beta_{\frac{\lambda}{D}}),
\end{eqnarray}
and
\begin{eqnarray}
	A_{D,ncp}(\alpha_{\frac{\lambda}{D}},\beta_{\frac{\lambda}{D}},\lambda) &=& i \left(g_{\alpha, 2}(\alpha_{\frac{\lambda}{D}}) - C_{x} f_{\alpha}(\alpha_{\frac{\lambda}{D}}) \right) \left( g_{\beta, 2}(\beta_{\frac{\lambda}{D}}) - C^{'}_{y}f_{\beta}(\beta_{\frac{\lambda}{D}}) \right) \nonumber \\
	&=& (4 \pi^{2} i a_{\lambda} b_{\lambda}) \left \{ \left(\frac{1}{(2 \pi i)^{2}} \frac{d^{2}}{d\alpha^{2}_{\frac{\lambda}{D}}} - \frac{1}{12}\right) f_{\alpha}(\alpha_{\frac{\lambda}{D}})\right\} \left \{ \left(\frac{1}{(2 \pi i)^{2}} \frac{d^{2}}{d \beta^{2}_{\frac{\lambda}{D}}} - \frac{1}{12} \right)f_{\beta}(\beta_{\frac{\lambda}{D}})\right \}.
\end{eqnarray}
$C_{x}$ (and $C^{'}_{y}$) was calculated, given that there is no pupil obscuration on the pupil plane:
\begin{eqnarray}
	\label{Cx}
	C_{x} &=& \tilde{m}(x) * P_{1}(x)\phi_{1,x}(x) \nonumber \\
	&=& \frac{w_{0}}{\xi_{x}}\mathrm{rect}\left(\frac{x}{w_{0}D}\right) * \left \{2\pi a_{\lambda} \left(\frac{x}{D}\right)^{2} P_{s,x}(x) \mathrm{rect}\left(\frac{x}{D}\right) \right\} \nonumber \\
	&=& \frac{2\pi a_{\lambda}}{12},
\end{eqnarray}
where $w_{0}$ was set to 2. Note that $C_{x}$ and $C^{'}_{y}$ have the same result under the condition that the pupil obscuration function is periodic. The contrast is defined as the ratio of the intensity distribution on the focal plane, $|A(\alpha_{\frac{\lambda}{D}},\beta_{\frac{\lambda}{D}})|^{2}$, to the peak of the ideal point-spread function, $|f(0, 0)|^{2}$. Therefore, the contrast distribution on the focal plane at the central wavelength, $C_{D,tot}(\alpha_{\frac{\lambda}{D}},\beta_{\frac{\lambda}{D}},\lambda)$, is expressed as 
\begin{eqnarray}
	\label{ideal_contrast}
	C_{D,tot}(\alpha_{\frac{\lambda}{D}},\beta_{\frac{\lambda}{D}},\lambda) &= & \frac{\left|A_{D,mask}(\alpha_{\frac{\lambda}{D}},\beta_{\frac{\lambda}{D}},\lambda=\lambda_{c})+A_{D,ncp}(\alpha_{\frac{\lambda}{D}},\beta_{\frac{\lambda}{D}},\lambda=\lambda_{c})\right|^{2}}{D^{2}} \nonumber \\
	&\simeq & C_{mask}(\alpha_{\frac{\lambda}{D}},\beta_{\frac{\lambda}{D}},\lambda=\lambda_{c}) + C_{ncp}(\alpha_{\frac{\lambda}{D}},\beta_{\frac{\lambda}{D}},\lambda) + C_{cross}(\alpha_{\frac{\lambda}{D}},\beta_{\frac{\lambda}{D}},\lambda),
\end{eqnarray}
where $C_{cross}(\alpha_{\frac{\lambda}{D}},\beta_{\frac{\lambda}{D}},\lambda)$ represents the cross-term of the two residual complex amplitudes, $C_{D,mask}(\alpha_{\frac{\lambda}{D}},\beta_{\frac{\lambda}{D}},\lambda=\lambda_{c})$ and $C_{D,ncp}(\alpha_{\frac{\lambda}{D}},\beta_{\frac{\lambda}{D}},\lambda)$, which are written as
\begin{eqnarray}
	C_{D,mask}(\alpha_{\frac{\lambda}{D}},\beta_{\frac{\lambda}{D}},\lambda=\lambda_{c}) &= & \left(\frac{U_{x}\left(B_{\frac{\lambda_{c}}{D}}, R\right) U_{y}\left(B_{\frac{\lambda_{c}}{D}}, R\right)}{4 \pi^{2} \alpha_{\frac{\lambda}{D}}\beta_{\frac{\lambda}{D}}} \right)^{2} \nonumber \\
	&\quad & \times \left\{\mathrm{sinc}(\pi \alpha_{\frac{\lambda}{D}}) - \cos(\pi \alpha_{\frac{\lambda}{D}}) \right\}^{2}  \left\{\mathrm{sinc}(\pi \beta_{\frac{\lambda}{D}}) - \cos(\pi \beta_{\frac{\lambda}{D}}) \right\}^{2},
\end{eqnarray}
and
\begin{eqnarray}
	C_{D,ncp}(\alpha_{\frac{\lambda}{D}},\beta_{\frac{\lambda}{D}},\lambda) &= & \left(\frac{a_{\lambda}b_{\lambda}}{\pi^{2} \alpha^{2}_{\frac{\lambda}{D}}\beta^{2}_{\frac{\lambda}{D}}}\right)^{2} \left\{\cos(\pi \alpha_{\frac{\lambda}{D}}) - \mathrm{sinc}(\pi \alpha_{\frac{\lambda}{D}}) + \frac{\alpha_{\frac{\lambda}{D}}}{2} \sin(\pi \alpha_{\frac{\lambda}{D}}) + \frac{\pi^{2}}{6}\mathrm{sinc}(\pi \alpha_{\frac{\lambda}{D}}) \right\}^{2}  \nonumber \\
	&\quad & \times \left\{\cos(\pi \beta_{\frac{\lambda}{D}}) - \mathrm{sinc}(\pi \beta_{\frac{\lambda}{D}}) + \frac{\beta_{\frac{\lambda}{D}}}{2} \sin(\pi \beta_{\frac{\lambda}{D}}) + \frac{\pi^{2}}{6}\mathrm{sinc}(\pi \alpha_{\frac{\lambda}{D}}) \right\}^{2}. 
\end{eqnarray}
Figures \ref{fig:ideal_limit} show the contrast curves limited by the stellar leak due to the new coronagraphic mask without a linear variable filter and the the high-order chromatic aberration generated in the ideal spectroscopic coronagraph without any alignment errors. The contrast is not limited by the high-order chromatic aberration but by the focal-plane mask optimized for this concept. The length of the mask along the spectral direction was set to 120 $\frac{\lambda}{D}$ at the central wavelength of 700 $nm$. When a linear variable filter is not applied to the focal plane, the light of the wavelength except for the central one does not pass through the center of the mask; the offset distances from the center of the mask are 46 and 52 $\frac{\lambda}{D}$ at 650 and 750 $nm$, respectively. Since the length of the mask in the unit of $\frac{\lambda}{D}$ is longer for the shorter wavelength, the contrast at 650 $nm$ is more limited than that at 750 $nm$. Thus, although the $10^{-10}$ contrast could be achieved at the inner working angle of $1 \frac{\lambda}{D}$, the observation bandwidth was limited to 100 $nm$, corresponding to that of 15\%. Note that, as discussed in Section \ref{subsec:mask}, this analytical estimation may provide the lower-limit on the observation bandwidth because the contrast may weakly depend on the length of the mask, $B_{\frac{\lambda}{D}}$, for the large $B_{\frac{\lambda}{D}}$. If a multilayer film working as a linear variable filter is applied to the substrate of the focal-plane mask, the length of the mask is optimized for each wavelength, and the light for all of the wavelengths passes through the center of the mask. As a result, the contrast curves over the observation bandwidth become the same as that of 700 $nm$ shown in the middle panel of Figure \ref{fig:ideal_limit}. 

The stellar leak due to the high-order chromatic aberrations is negligible because the ideal Offner-type imaing spectrograph without any alignment errors does not generate a non-axis aberration; the second-order terms of the aberration function limit the contrast curves. Furthermore, the stellar leak could be perfectly suppressed at the central wavelength of 700 $nm$, due to the optimized Offner-type spectrograph. Note, however, that the alignment errors of the optical system degrade the performance of this system, which will be discussed in the following subsection; this performance highlights the principal limit of the spectroscopic coronagraph system.

\subsection{Application to large segmented telescopes} \label{performance_on_actual_pupils}
\begin{figure}
	 \centering
	\includegraphics[scale=0.6,height=15cm,clip]{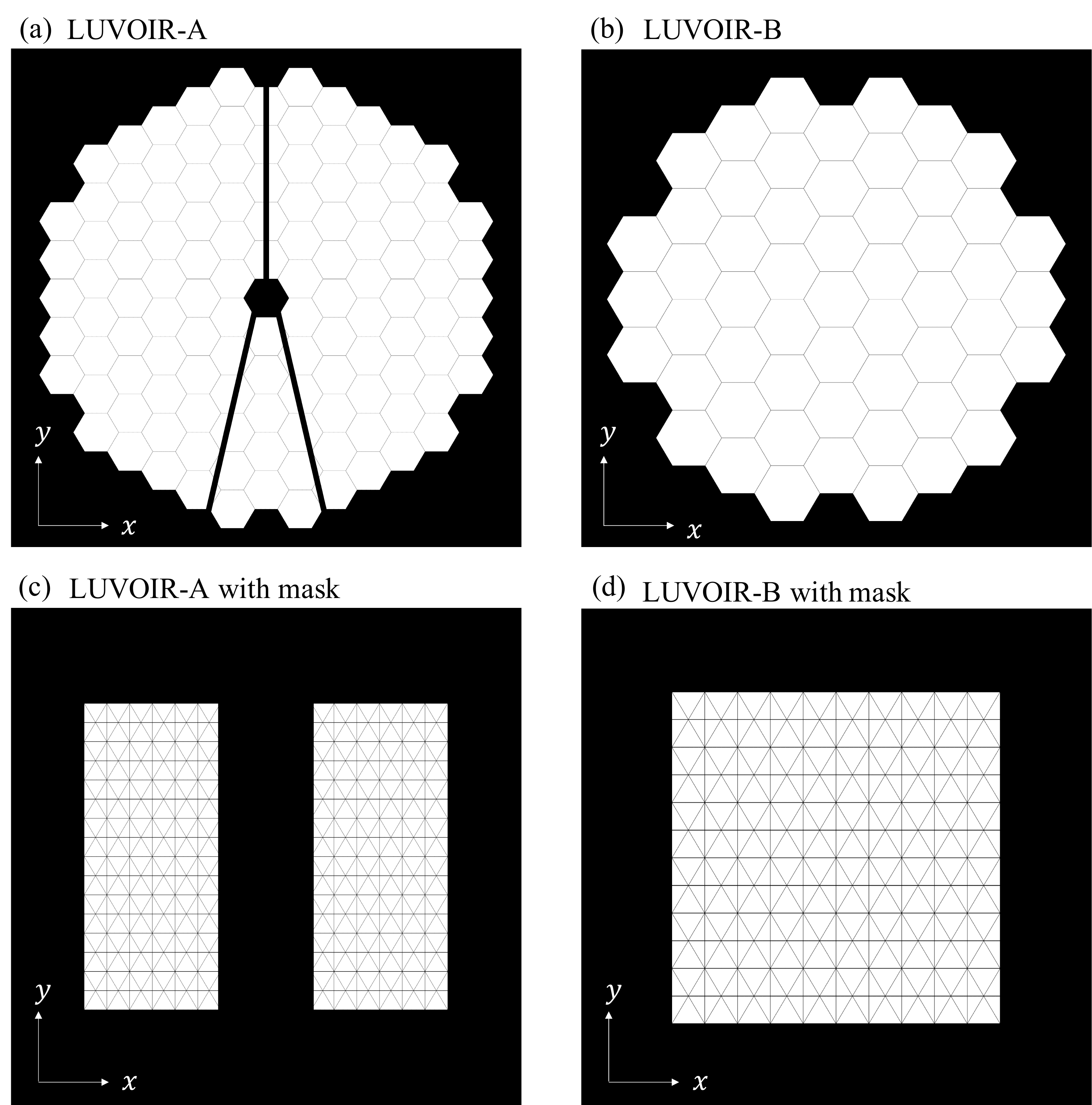}
	\caption{(a) Pupils of LUVOIR-A (a) and -B (b) and pupils shielded by masks optimized for LUVOIR-A (c) and -B (d). The pupils of the LUVOIR-A and -B were produced based on Figure 8-10 of the LUVOIR final report \citep{LUVOIR+2019}.}
	\label{fig:LUVOIR_pupil}
\end{figure}

We applied the analytical expressions derived in the previous sections to the LUVOIR telescope designs and derived the expected performance of this coronagraph concept, as shown in this subsection. Panels (a) and (b) of Figures \ref{fig:LUVOIR_pupil} show the pupils of LUVOIR-A and -B with diameters of 15 and 8 m, respectively, according to the LUVOIR final report \citep{LUVOIR+2019}, respectively. The width of the gap was set to 0.1 \%. There are various types of pupils for the LUVOIR telescope design, and the performance of this coronagraph concept depends on the design. Panels (c) and (d) of Figure \ref{fig:LUVOIR_pupil} show the pupils shielded by the optimized masks for LUVOIR-A and -B, respectively. $\xi$ shown in Equation \ref{xi} is constant along both the $x$ and $y$ axes. As a result, the region overlapped by the secondary mirror and spiders of the LUVOIR-A concept was blocked by a bar-like mask with a width of 4.2 $m$. The mask for LUVOIR-A had a relatively large impact on the throughput of the coronagraph. \deleted{The throughputs of the masks optimized for LUVOIR-A and -B are 51.1 \% and 69.4 \%, respectively.} \added{The throughput efficiencies of the entrance pupils optimized for LUVOIR-A and -B, $\eta_{p}$, are 0.511 and 0.694, respectively.} The lengths ($y$) and widths ($x$) of the masked pupils for LUVOIR-A and -B are 9.8 x 11.3 and 5.0 x 5.7 $m$, respectively. Since the sizes of the masked pupils are reduced from the original ones, the inner working angles of LUVOIR-A and -B increase by a factor of approximately 1.3 - 1.5. 

\begin{figure}
	 \centering
	\includegraphics[scale=0.6,height=12cm,clip]{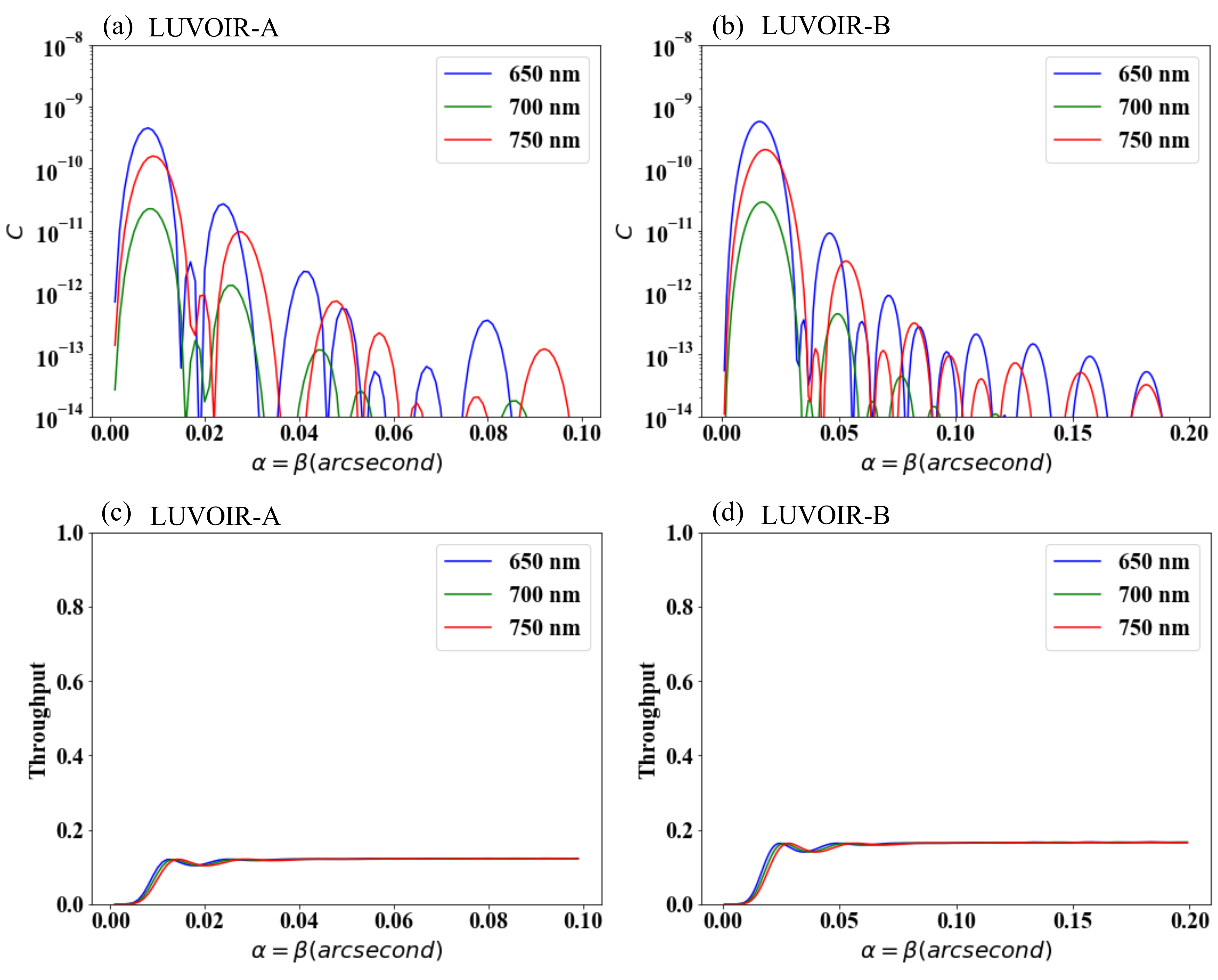}
	\caption{(a) Contrast curves limited by the coronagraph masks optimized for LUVOIR-A (a) and -B (b) as a function of the angular separation from the host stars at three wavelengths, 650, 700, and 750 $nm$, and the throughputs of the coronagraph masks optimized for LUVOIR-A (c) and -B (d) as a function of the angular separation from the host stars. }
	\label{fig:LUVOIR_contrast_throughput}
\end{figure}

Based on the masked pupils, we calculated the contrast curves for the optimized spectroscopic coronagraph as a function of the angular separation from the host star at three wavelengths, 650, 700, and 750 $nm$ (Panels of (a) and (b) of Figure \ref{fig:LUVOIR_contrast_throughput}). The contrast of $10^{-10}$ could be achieved at 12 and 28 $milli-arcsecond$ ($mas$) over the wavelength range of 650 to 750 $nm$. Panels (c) and (d) of Figure \ref{fig:LUVOIR_contrast_throughput} show the throughputs of only the spectroscopic coronagraphs for LUVOIR-A and -B, respectively. The throughputs of the off-axis sources beyond 12 and 28 $mas$ are 0.12 and 0.19 for LUVOIR-A and -B, respectively. \deleted{The total throughput of the coronagraph is equal to a muliplication of the efficiency of the entrance pupil and the throughput for off-axis sources, corresponding to $T_{m}$ shown in Equation \ref{M_alpha}.} \added{The total throughput of this coronagraph concept is written as $\eta_{p} (T_{m})^{i}$, where $(T_{m})^{i}$ shown in Equation \ref{M_alpha} represents the throughput efficiency of the focal-plane mask for the $i$-th order null.} $T_{m}$ is 0.7 for both LUVOIR-A and -B, given that the width of the gap is negligible (i.e., $\xi_{i} = 0$). We note that the modulation functions of both the first- and second-stages for  LUVOIR-A are set to be parallel to along the $y$-axis (i.e., $\beta$-axis), considering that the entrance pupil optimized for LUVOIR-A has a bar-like obscuration; if the two spectroscopic coronagraphs place in succession, orthogonal to each other, $T_{m}$ of the modulation function along the $x$-axis decreases down to 0.5 because $\xi_{x}$ is 0.7.

Based on these considerations, this spectroscopic coronagraph concept works at very small inner working angles of 12 and 28 $mas$ at 750 $nm$, corresponding to 1.2 and 1.5 $\frac{\lambda}{D}$, for LUVOIR-A and -B in terms of the contrast and throughput of the off-axis sources, respectively. We note, however, that the telescope pointing jitter should be fully suppressed because of the fourth-order null of this coronagraph; the stellar leak will dominate the planet light at the inner working angle if the pointing jitter is larger than 0.01 $\frac{\lambda}{D}$. \deleted{Conversely, while the inner working angles of the current baseline coronagraphs for LUVOIR-A and -B are 3.7 and 2.5 $\frac{\lambda}{D}$ \citep{Stark+2019}, which are larger than that of this coronagraph concept, it are relatively insensitive to the low-order aberrations.} \added{Conversely, while the inner working angles of the baseline LUVOIR-A and -B coronagraph designs are larger (3.7 and 2.5 $\lambda$; \cite{Stark+2019}), they are relatively insensitive to low-order aberrations.} In addition, the throughput of the vector vortex coronagraph as the baseline of LUVOIR-B is approximately 40 $\%$ beyond 7 $\frac{\lambda}{D}$, which is two times higher than that of this coronagraph. Thus, although the baseline coronagraphs outperform the spectroscopic coronagraph in terms of the robustness and throughput, this coronagraph concept potentially improves the inner working angles of LUVOIR-A and -B, and the large space telescope could characterize habitable planet candidates not only around G- and K-type stars beyond 20 $pc$ but also around nearby M-type stars. Thus, this concept is complementary to the baseline coronagraphs of LUVOIR-A and -B. 

\section{Tolerance analysis} \label{sec:torelance}
This section focuses on the tolerance analysis related to propagations of the following four errors in the spectroscopic coronagraph: (1) low-order aberrations, (2) alignment error of the spectroscopic coronagraph, (3) in-plane non-uniformity of the diffraction grating, and (4) production error of the spherical mirror. We discuss how the four unconsidered factors degraded the spectroscopic coronagraph performance. 

\subsection{Sensitivity to low-order aberrations}
We have not discussed the impact of a low-order \replaced{aberration}{aberrations} occured due to a telescope pointing jitter and finite stellar disk on the achievable contrast so far. The originally proposed coronaraphic mask was designed such that the complex amplitude is modulated along only one direction. Combining two coronagraph systems in succession to be paralell or orthogonal to each other achieves a fourth-order null, which can result in high-contrast imaging at 1 $\frac{\lambda}{D}$ for large telescopes when the pointing jitter is less than 0.01 $\frac{\lambda}{D}$. In fact, only the cross-term of $x$- and $y$-tilts (i.e., second-order term) propagates through the final Lyot stop because the tilt error can be written as an $x$- or $y$-dependent function. According to Equation \ref{iedal_2L}, the complex amplitude on the Lyot stop is 
\begin{equation}
	A_{2L}(x,y,\lambda) = - P_{1}(x)P_{1}(y)\phi_{1,x}(x)\phi_{1,y}(y),
\end{equation}
where $C_{x}$ and $C_{y}^{'}$ in Equation \ref{iedal_2L} are 0 for the case of the pure tip-tilt aberration. Using Equation \ref{phi_x_y}, the above Equation is rewritten as
\begin{equation}
	\label{A_2L_tilt}
	A_{2L}(x,y,\lambda) = - (4 \pi^{2} a_{\lambda} b_{\lambda})P_{1}(x)P_{1}(y) \left( \frac{x}{D} \right) \left( \frac{y}{D} \right).
\end{equation}
The stellar leakage \deleted{occurred} due to the low-order \replaced{aberration}{aberrations} on the detector plane is suppressed with a fourth-order null, corresponding to the square of the second-order term. However, the imaging of habitable planet candidates around G- and K-type stars requires 0.01 $\frac{\lambda}{D}$ or less of the telescope pointing jitter because the fourth-order coronagraph is highly sensitive to low-order aberration compared to higher-order coronagraphs, such as a vector vortex coronagraph with a topological charge higher than 6 \citep{Mawet+2010, Ruane+2018} and an eighth-order band-limited mask \citep{Kuchner+2005}.

On the other hand, this new coronagraphic mask optimized for the spectroscopic coronagraph concept modulates the complex amplitude along both the two axes of the focal plane, which generates additional stellar leakage even for non-aberrated wavefront, as discussed in Section \ref{subsec:mask}. We evaluate the impact of the optimized coronagraphic mask on the achievable contrast. If only the tilt errors along the $x$ and $y$ axes exist on the entrance pupil for simplicity, Equation \ref{A_res_pupil} becomes
\begin{eqnarray}
	A_{2L,mask}(x,y,\lambda = \lambda_{c}) &=& P_{1}(x)P_{1}(y) \left\{\Delta \tilde{m}(x,y,\lambda=\lambda_{c}) * P_{1}(x)P_{1}(y)\left(1 + 2\pi i a_{\lambda} \left( \frac{x}{D} \right) + 2\pi i b_{\lambda} \left( \frac{y}{D} \right) \right) \right\} \nonumber \\        
		           &=& \frac{1}{2 R} \left(\frac{\lambda_{c}}{D} \right) \left(\frac{x}{D} \right) P_{1}(x)P_{1}(y)B_{\frac{\lambda_{c}}{D}} \left(U_{y} + \delta_{a_{\lambda}} + \delta_{b_{\lambda}} \right),               
\end{eqnarray}
where $U_{y}$ is almost constant (see Section \ref{subsec:mask}), and $\delta_{a_{\lambda}}$ and $\delta_{b_{\lambda}}$ represent the additional stellar leak \replaced{occurred}{occurring} due to the two-dimensional dependence of the coronagraphic mask in the presence of the low-order aberration. Since $a_{\lambda}$ and $b_{\lambda}$ are much smaller than $\frac{1}{2\pi}$ for the general high-contrast instruments, the cross-terms of the dependency of the foca-plane mask on the two axes and the low-order aberrations, $\delta_{a_{\lambda}}$ and $\delta_{b_{\lambda}}$, are much smaller than $U_{y}$; the additional stellar leak originated from $\delta_{a_{\lambda}}$ and $\delta_{b_{\lambda}}$ is negligible compared to that of the original coronagraphic mask in the presence of the low-order aberration shown in Equation \ref{A_2L_tilt}. 

Thus, this spectroscopic coronagraph has the same sensitivity to the low-order aberrations as the original coronagraphic mask that modulates the complex amplitude along only one direction. 

\subsection{Impact of alignment error on the contrast}
We assumed, as described in the previous section, that no alignment error was present in the spectroscopic coronagraph. However, the alignment error generates high-order chromatic aberrations even if the wavefront at the central wavelength is perfectly compensated. We evaluate the high-order chromatic aberration generated by the spectroscopic coronagraph with alignment errors and derive the achievable contrast on the detector plane. As shown in Figure \ref{fig:optical_design}, the three-dimensional coordinate system, $x-y-z$, is defined, and the angle of rotation around the $i$-axis is set to $\theta_{i}$. The alignment errors of the reflection grating and spherical mirror have six degrees of freedom. We considered eleven degrees of freedom in total for the alignment errors of this system. Note that the angle of rotation around the $z$-axis is not considered because the spherical mirror is symmetrical around the $z$-axis. The alignment errors for this calculation are set to $\pm 20$ $\mu m$ and $\pm 0.005$ $deg$ as the achievable accuracy without any sophisticated alignment method \citep[e.g.,][]{Winrow+2011}. Since the optical axis of the reflection grating is matched to that of the spherical mirror in the ideal case, it is preferable to mount the grating and spherical mirror with a same structure. The three-dimensional relative position of the two surfaces could be accurately measured with a three-dimensional measuring machine under the condition that the two optical elements are mounted with the same structure. Note that $\pm 0.005$ $deg$ corresponds to approximately $\pm20$ $\mu m$ displacement of the grating or spherical mirror relative the distance between the spherical mirror and reflection grating. 

Tables \ref{tab:aberration_grating} and \ref{tab:aberration_mirror} show the defocus, astigmatism (0/90), and astigmatism (45/135) at the central wavelength of 700 $nm$ for the alignment errors of the reflection grating and spherical mirror, respectively. Note that higher order aberrations than defocus and astigmatism, such as coma and trefoil, are smaller than 0.1 $nm$. The astigmatism (45/135), which is the cross-term of the $x$- and $y$-dependent aberration functions, is newly generated by their displacements along the $x$-axis and rotations around $\theta_{y}$. This astigmatism (45/135) is generated by shifting the beam along the $x$-axis, and it has a larger impact on the coronagraph performance than on-axis aberrations because the astigmatism (45/135) is the first-order term of the aberration function (see Equation \ref{phi}).  

\begin{figure}
	 \centering
	\includegraphics[scale=0.6,height=7cm,clip]{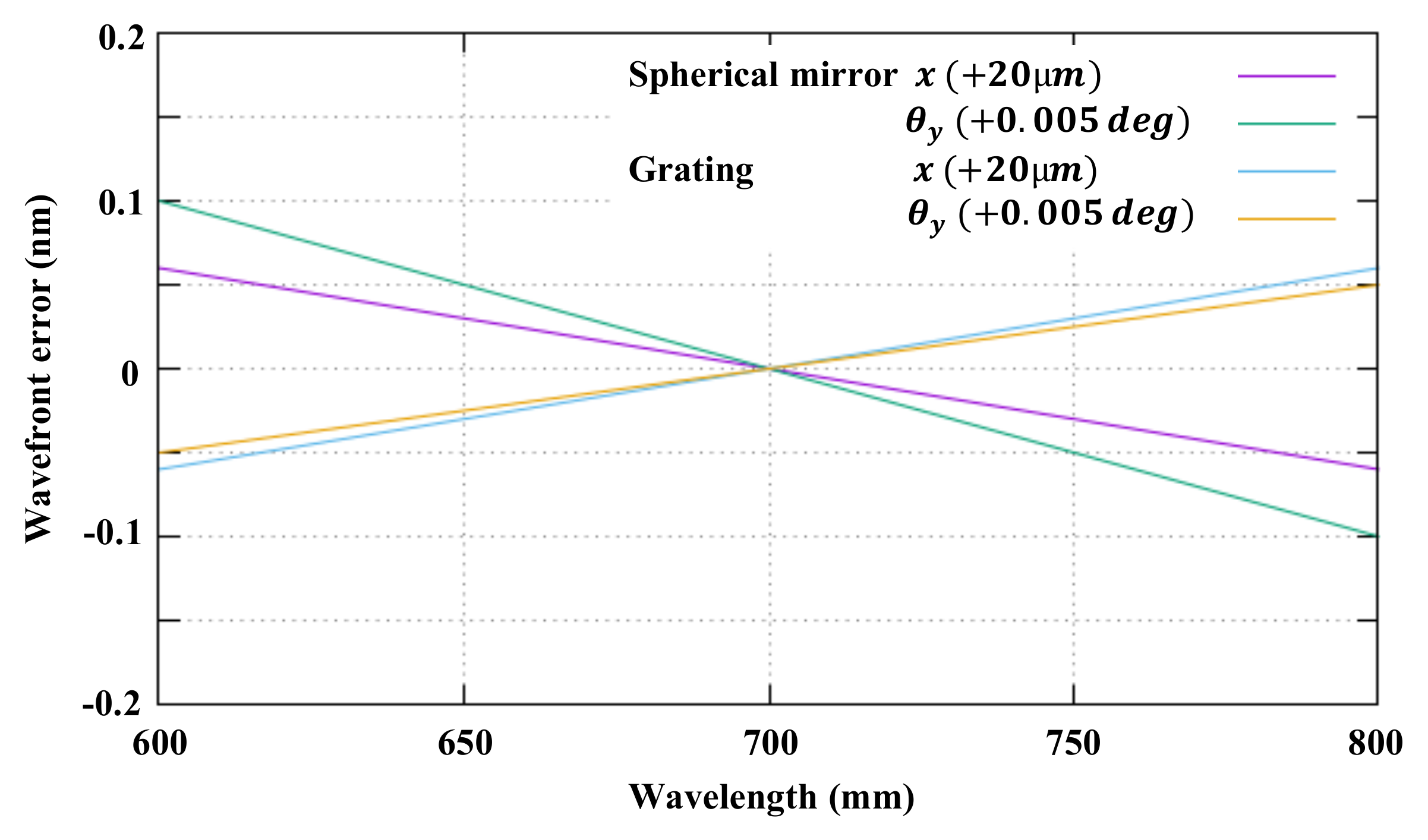}
	\caption{High-order chromatic astigmatism (45/135) in the wavelength range of 600 to 800 $nm$. The astigmatism (45/135) for each alignment error is set to 0 at 700 $nm$.}
	\label{fig:chromatic_aberration}
\end{figure}

\begin{figure}
	 \centering
	\includegraphics[scale=0.6,height=7cm,clip]{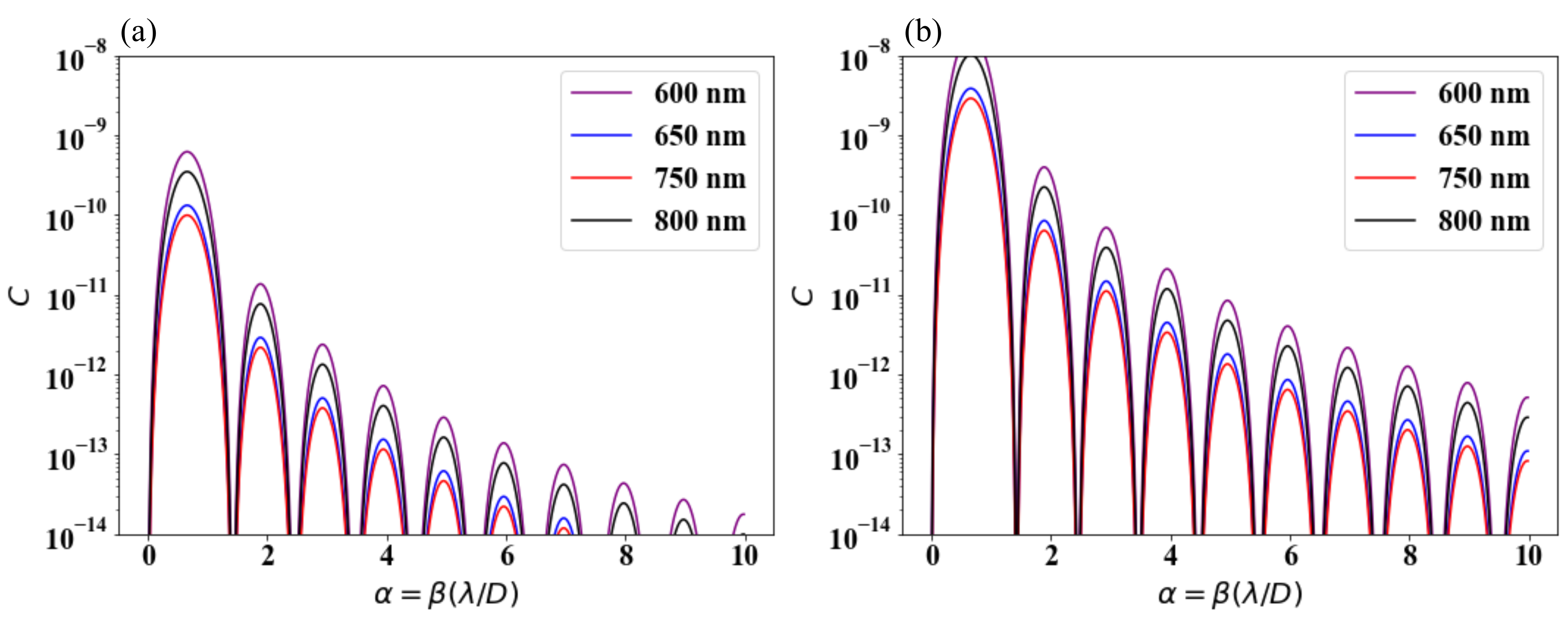}
	\caption{Contrast curves along $\alpha = \beta$ for the fiducial (left) and worst cases (right) of the optical alignment errors.}
	\label{fig:contrast_chromatic_aberration}
\end{figure}

We, hereinafter, focus on the impact of the astigmatism (45/135) on the contrast on the detector plane. Since the aberration function, $\phi (x,y)$, is written as $2\pi a^{'}_{\lambda} \left (\frac{x}{D} \right) \left( \frac{y}{D} \right)$, the contrast curve on the focal plane becomes
\begin{equation}
	\label{alignment_error_contrast}
	C_{D,ncp}(\alpha_{\frac{\lambda}{D}},\beta_{\frac{\lambda}{D}},\lambda) = \left(\frac{a^{'}_{\lambda}}{2 \pi \alpha^{2}_{\frac{\lambda}{D}}\beta^{2}_{\frac{\lambda}{D}}}\right)^{2} \left\{\mathrm{sinc}(\pi \alpha_{\frac{\lambda}{D}}) - \cos(\pi \alpha_{\frac{\lambda}{D}}) \right\}^{2} \left\{\mathrm{sinc}(\pi \beta_{\frac{\lambda}{D}}) - \cos(\pi \beta_{\frac{\lambda}{D}}) \right\}^{2}. 
\end{equation}
Note that $C_{xy}$ and $C_{xy}^{'}$ in Equation \ref{complex_amplitude_2L_result} become 0 because the aberration function, $\phi(x,y)$, is odd at each axis. Given that the aberration at the central wavelength of 700 $nm$ is corrected by deformable mirrors before the coronagraph system, the difference between the aberrations at 700 $nm$ and the other wavelength (i.e., high-order chromatic aberration) directly limits the coronagraph performance. Figure \ref{fig:chromatic_aberration} shows the residual astigmatism (45/135) over the wavelength range of 600 to 800 $nm$. Since the spherical mirror and reflection grating face each other, the displacements and rotations of the grating and spherical mirror along the same direction reduce the astigmatism (45/135). Here, because the alignment error range is within the alignment accuracy in general, we consider two cases as the alignment errors: (a) fiducial and (b) worst cases. For the fiducial case, all the alignment errors apply the maximum displacement along the $x$-axis and the maximum rotation along the $y$-axis, $\theta_{y}$, and the directions for all of the errors are same; the astigmatism (45/135) will be canceled to some extent. For the worst case, the alignment errors apply the maximum values, and their directions are determined so that all the astigmatism (45/135) aberrations due to the alignment errors are accumulated. Figure \ref{fig:contrast_chromatic_aberration} shows the contrast curves along $\alpha = \beta$ for the two cases. While the contrast of $10^{-10}$ is almost achieved  at 1 $\frac{\lambda}{D}$ over the wavelength range, the $10^{-10}$ contrast is achieved beyond 2 $\frac{\lambda}{D}$ for the worst case. Note that, because the alignment errors set for this calculation could be reduced, the contrast would be considerably improved at an inner working angle. 

Thus, the Offner-type imaging spectrograph applied in the spectroscopic coronagraph concept could minimize the impact of high-order chromatic aberrations on the performance. Combining the coronagraph applying one-dimensional modulation mask with the Offner-type imaging spectrograph enlarges the observation wavelength range.

\begin{table}[htb]
	\begin{center}		
		\caption{Wavefront aberration at 700 $nm$ for the alignment error of the reflection grating in the unit of $nm$}
		\scalebox{0.8}{
 		\begin{tabular}{| l | c | c | c | c | c | c |} \hline 
    		 & $x$ (+20$\mu m$) & $y$ (+20$\mu m$) & $z$  (+20$\mu m$) & $\theta_{x}$ (+$0.005 deg$) & $\theta_{y}$ (+$0.005 deg$) & $\theta_{z}$ (+$0.005 deg$)  \\ \hline \hline
   		Defocus & 0.00  & 0.00 & 0.21 & 10.99 & 0.00 & 0.00 \\ \hline
    		Astigmatism (0/90) & 0.00 & -0.35 & 0.14 & 0.35 & 0.00 & 0.00 \\ \hline
		Astigmatism (45/135) & 0.42 & 0.00 & 0.00 & 0.00 & 0.35 & 0.00 \\ \hline
  \end{tabular}
  }
  \label{tab:aberration_grating}
  \end{center}
\end{table}

\begin{table}[htb]
	\begin{center}
		\caption{Wavefront aberration at 700 $nm$ for the alignment error of the spherical mirror in the unit of $nm$}
		\scalebox{0.8}{
 		\begin{tabular}{| l | c | c | c | c| c |} \hline 
    		 & $x$ (+20$\mu m$) & $y$ (+20$\mu m$) & $z$  (+20$\mu m$) & $\theta_{x}$ (+$0.005 deg$) & $\theta_{y}$ (+$0.005 deg$)  \\ \hline \hline
   		Defocus & 0.00 & -0.21 & -22.12 & 0.35  & 0.00 \\ \hline
    		Astigmatism (0/90) & 0.00  & 0.42 & -0.14 & -0.63 & 0.00 \\ \hline
		Astigmatism (45/135) & -0.42 & 0.00 & 0.00 & 0.00 & -0.70 \\ \hline
  \end{tabular}
   }
  \label{tab:aberration_mirror}
  \end{center}
\end{table}

\subsection{In-plane non-uniformity of the diffraction efficiency}
The in-plane non-uniformity of the grating efficiency exists because the Offner-type spectrograph applies a convex reflection grating. As shown in Appendix \ref{sec:appendix_b}, the in-plane non-uniformity has a weak linear dependence of the axis, along which the spectrum forms. Therefore, the in-plane non-uniformity of grating efficiency in the first-stage coronagraph can be described by the following Equation: 
\begin{equation}
	\eta(x,y) = 1 - \left \{ G + H \left(\frac{y}{D} \right) \right \},
\end{equation}
where $\eta(x,y)$ ranges from 0 to 1, and the coefficients, $G$ and $H$, are constant. Table \ref{tab:pupil_efficiency} shows the coefficients, $G$ and $H$, for the optical design shown in Section \ref{subsec:optical_design}. They were calculated from the wavefront map acquired by the optical simulations. Since the dependency of this in-plane non-uniformity is perpendicular to the coronagraph mask's modulation direction, the stellar leak caused by this in-plane non-uniformity does not propagate through the Lyot stop. Thus, this in-plane non-uniformity does not degrade the spectroscopic coronagraph performance at all.   

\begin{table}[htb]
	\begin{center}
		\caption{Wavelength dependency of the in-plane non-uniformity of the diffraction efficiency}
  		\begin{tabular}{| l | c | c | c | c| c |} \hline 
    		Item & 600 $nm$ & 650 $nm$ & 700 $nm$ & 750 $nm$ & 800 $nm$ \\ \hline \hline
    		Coefficient G & 1.63E-02  & 2.41E-03 & 3.87E-04 & 6.23E-03 & 1.68E-02 \\ \hline
    		Coefficient H & 2.25E-05 & 3.28E-05 & 3.70E-05 & 4.00E-05 & 4.23E-05 \\ \hline
  \end{tabular}
  \label{tab:pupil_efficiency}
  \end{center}
\end{table}

\subsection{Structure function of the spherical mirror}
Thus far, we have investigated the spectroscopic coronagraphy concept, assuming that the production errors of the optical elements are composed of the Offner-type spectrograph do not exist. However, the figures of the convex reflection grating and the spherical mirror differ from the ideal ones. Since the convex reflection grating disperses the white light, the production error of the convex grating can be corrected over the observation bandwidth by the deformable mirrors in the upstream section of the coronagraph system. In contrast, the production error of the spherical mirror limits the performance of the spectroscopic coronagraph because the beam position on the spherical mirror differs depending on the wavelength. Conversely, only a portion of the spherical mirror is used for this spectroscopic coronagraph system; the area reflecting the beam is approximately one-hundredth of the area of the spherical mirror. The difference in the beam positions on the spherical mirror over the observation bandwidth are within one-tenth of the beam diameter. In other words, the spherical mirror's structure-function at the scale of one-hundredth of the mirror diameter degrades the coronagraph performance. Given that the wavefront at the central wavelength of 700 $nm$ is perfectly corrected, the residual wavefront error in the other wavelengths is written as 
\begin{eqnarray}
	\Delta \phi(x,y) &=& \phi(x,y +\Delta y) - \phi(x,y)  \nonumber \\
	&\simeq& \Delta y \left. \frac{d\phi(x, y)}{dy} \right|_{\Delta y = 0},   
\end{eqnarray}
where $\Delta y$ is the difference of the beams' positions on the spherical mirror between the central wavelength and other wavelengths, and the white light is assumed to be dispersed along the $y$-axis, as shown in Figure \ref{fig:optical_design}. 

\begin{figure}
	 \centering
	\includegraphics[scale=0.6,height=4.5cm,clip]{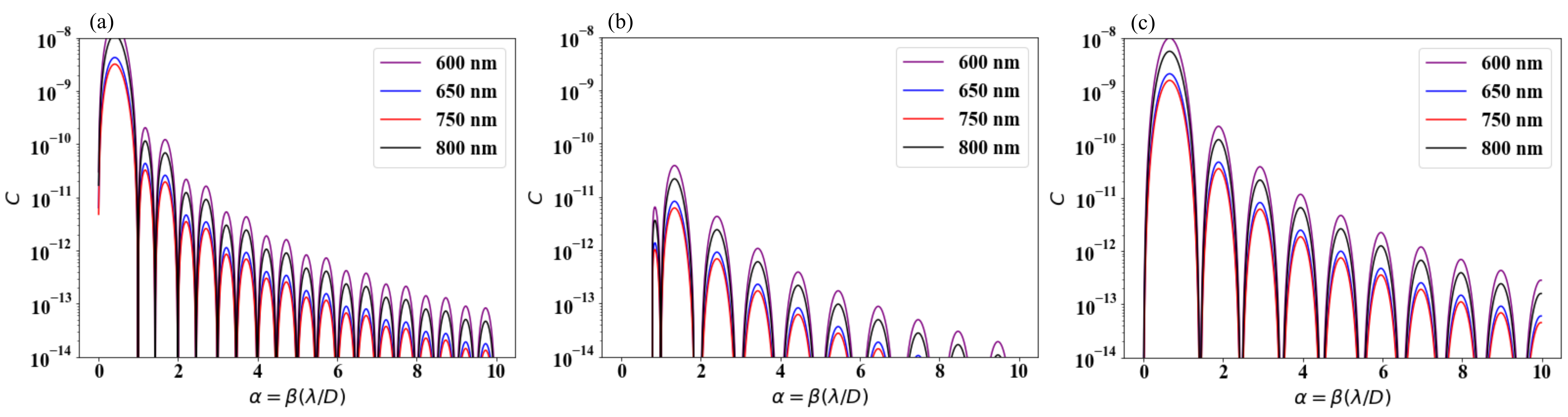}
	\caption{Contrast curves along $\alpha = \beta$ for pupils having aberrations of (a) $xy$, (b) $x^{2}y$, and (c) $xy^{2}$, generated by the manufacturing errors of the spherical mirror. We assumed that the wavefront error at the central wavelength of 700 $nm$ is perfectly corrected. The difference of the beams' positions on the spherical mirror between the central wavelength and 650 (600) or 750 (800) $nm$ was set to one-twentieth (one-tenth) of the beam diameter. }
	\label{fig:manufacture_error}
\end{figure}

As mentioned in Section \ref{subsec:propagation}, the cross-term of the aberration function, $\phi (x) \phi(y)$, limits the contrast primarily. Furthermore, the intensity distribution on the focal plane is proportional to $\left(\alpha_{\frac{\lambda}{D}}\right)^{-n}$ and $\left(\beta_{\frac{\lambda}{D}}\right)^{-n}$ for a pupil with the $n$-th order aberration function; as the $n$-th order increases, the contrast goes to 0 more quickly in the outer region of the focal plane. We focus on the cross-terms included in the third- or lower order Zernike polynomial function. We evaluate the contrast curve based on the analytical expressions in Section \ref{subsec:performance}. Figure \ref{fig:manufacture_error} shows the contrast curve at 600, 650, 750, and 800 $nm$ under the assumption that the reflection wavefront after the spherical mirror has a 1 $nm$ phase error due to the production error.  $\Delta y$ was set to one-twentieth of the beam diameter for 650 and 750 $nm$ and a tenth for 600 and 800 $nm$, based on the designed optical system. As shown in Figure \ref{fig:manufacture_error}, the contrast curves do not largely change with the type of cross-term function, and the $10^{-10}$ contrast can be achieved close to the inner working angle of 1 $\frac{\lambda}{D}$. Focusing on the relationship between the Zernike polynomial function and the mirror's structure-function \citep{Hvisc+2017}, the wavefront aberration is approximately proportional to the spatial scale of the mirror. Therefore, the shape accuracy of the spherical mirror should be less than 5 $nm$ for the high-contrast imaging of terrestrial planets orbiting Sun-like stars.  

\section{Conclusion}
We proposed a new approach for characterizing nearby terrestrial candidates over a wide observation bandwidth: we develop a coronagraph that enables high-contrast imaging at 1 $\frac{\lambda}{D}$ on segmented telescopes \citep{Itoh+2020}. This approach was named "spectroscopic fourth-order coronagraphy." Focusing on the fact that the original complex mask modulates the complex amplitudes of astronomical objects along one direction on the focal plane, we combined two spectrographs with a coronagraph system having an optimized focal-plane mask for this approach to enlarge the observation bandwidth of the high-contrast system. While the modulation period of the focal-plane mask is optimized for each spectral element resolved by the spectrograph, the new mask slightly differs from the original one-dimensional modulation function; the modulation period changes along the spectral direction. Also, since the white light is dispersed by a diffraction element, the optical path between the diffraction element and the focal plane changes depending on the wavelength; the non-common path error (i.e., high-order chromatic aberration) occurs before the light transmits through the focal-plane mask. 

We analytically investigated the newly generated system based on the following two points: (1) the impact of the newly optimized focal-plane complex on the coronagraphic performance and (2) how the high-order chromatic aberration propagates through the fourth-order coronagraph system. We found that the optimized focal-plane mask introduces an aberration equivalent to a tilt error along its modulation direction because the modulation period lineary changes along the spectral direction. As a result, the stellar leak is proportional to the fourth-power of the length of the mask for the fourth-order coronagraph system. The length of the mask should be optimized according to the target contrast of each instrument. Note, however, that the strong dependence of the length of the mask on the contrast should be carefully treated for a large focal-plane mask (i.e., large $B_{\frac{\lambda}{D}}$) because the coronagraphic mask is not analytically approximated well for the large $B_{\frac{\lambda}{D}}$; the dependence may be weak for the large $B_{\frac{\lambda}{D}}$. The stellar leak also increases as the focal point of the light further deviates from the center of the mask; the contrast becomes worse at the edge of the observation bandwidth. Regarding the latter point, we noticed that only one-axis dependent aberration functions do not transmit through the coronagraph system with the fourth-order null; further, the cross-terms of these functions limit the performance of the coronagraph. 

Based on these analytical considerations, we designed a spectroscopic coronagraph with an Offner-type spectrograph that does not generate non-axis aberration but the defocus and astigmatism (0/90) abrrations. The observation band ranges from 600 to 800 $nm$, and the resolving power of the spectrograph is about 670. We analytically derived the contrast curves on the detector plane from the residual complex amplitude on the Lyot plane, which propagates through the fourth-order spectroscopic coronagraph. We noticed that the length of the mask should be limited to 120 $\frac{\lambda}{D}$ at the central wavelength of 700 $nm$ for achieving the $10^{-10}$ contrast at the inner working angle of 1 $\frac{\lambda}{D}$; the wavelength ranges from 650 to 750 $nm$, corresponding to the bandwidth of 15 \%. We applied the analytical expression to the LUVOIR telescope designs, and we derived the contrast curves and throughputs for the coronagraph masks optimized for the LUVOIR-A and -B. This coronagraph concept works well at the angular separation of 12 and 28 $mas$ at 750 $nm$, corresponding to 1.2 and 1.5 $\frac{\lambda}{D}$, for the LUVOIR-A and -B in terms of the contrast and throughput, respectively. However, we should note that low-order aberration, such as the telescope pointing jitter, should be suppressed down to 0.01 $\frac{\lambda}{D}$ for achieving the high contrast of $10^{-10}$ at the small inner working angle. Conversely, while the baseline coronagraphs for LUVOIR-A and -B, which are the Apodized Pupil Lyot Coronagraph and Vector vortex coronagraph with a support of a deformable mirror, are more robust agiainst the telescope pointing jitter, the inner working angles for the LUVOIR-A and -B are 3.7 and 2.5 $\frac{\lambda}{D}$, respectively. Thus, this spectroscopic coronagraph concept is complementary to the baseline coronagraphs.

Finally, we performed torelance analysis related to propagations of the following four factors through this spectroscopic coronagraph: (1) alignment error of the spectroscopic coronagraph, (2) the low-order aberration, (3) the in-plane non-uniformity of the diffraction efficiency, and (4) the production errors of the spherical mirror. Regarding the first factor, we numerically simulated the wavefront aberrations for the spectroscopic coronagraph design with the alignment errors of the optical elements and derived the contrast curves on the focal plane by substituting the aberrations to the analytical expressions. Note that the alignment errors for only the convex grating and the spherical mirror were considered because the other optical elements put in the common path. We found that the cross-term of the $x$- and $y$-dependent functions, astigmatism (45/135), generates in the optical system with the alignment errors. However, the $10^{-10}$ contrast could still be achieved at 1-2 $\frac{\lambda}{D}$ under the condition that the displacement and offset angle of the optical elements are within $\pm 20$ $\mu m$ and $\pm 0.005$ degree, respectively. Regarding the second factor, the newly optimized coronagraphic mask has the same sensitivity to the low-order aberration as the originally-proposed one-dimensional amplitude mask; this coronagraph concept achieves the fourth-order null. Furthermore, since the in-plane efficiency of reflective grating weakly depends on only the spectral direction, the in-plane non-uniformity does not affect the coronagraphic performance at all. Conversely, the cross-term of aberrations due to the production error of the spherical mirror could limit the contrast of this concept. The manufacturing error of the spherical mirror must be within $\pm$ 5 $nm$ to achieve the $10^{-10}$ contrast at the inner working angle of 1 $\frac{\lambda}{D}$.

Thus, using this approach, the spectral characterization of nearby habitable planets can be performed using future large-space telescopes and ELTs. As the next step, we will perform numerical simulations to accuratelly estimate the observation bandwidth through investigating the dependence of the length of the mask on the stellar leak for the large $B_{\frac{\lambda}{D}}$. Since the complex amplitude of the stellar light at the large $B_{\frac{\lambda}{D}}$ ($> 100\frac{\lambda}{D}$) is much smaller than its peak, the modulation function gives a negligible impact on the complex amplitude at that region. The observation bandwidth may be increased without the linear variable filter on the focal plane if the dependence of the length of the mask on the stellar leak is weak at the large $B_{\frac{\lambda}{D}}$. 

\acknowledgments

We appreciate Dr. Toru Yamada and Dr. Takahiro Sumi for useful discussions and valuable comments on this spectroscopic coronagraphy concept. We also would like to express our sincere gratitude to anonymous referee for evaluating this study and providing a number of valuable suggestions. TM is supported by Grand-in-Aid from MEXT of Japan, No. 19H00700.

\appendix
\section{Taylor Expansion of Modulation function}
\label{sec:appendix_a}
When $\chi \beta_{\frac{\lambda}{D}}$ is not considered to be fully smaller than $\alpha_{\frac{\lambda}{D}}$, the modulation function should be expanded by high-order Taylor series. The second-order Taylor expansion of the modulation function is  
\begin{eqnarray}
	\label{m_spe_approx_appendix}
	m_{spectrum}(\alpha_{\frac{\lambda}{D}},\beta_{\frac{\lambda}{D}}) &\simeq & m(\alpha_{\frac{\lambda}{D}}) + \chi \beta_{\frac{\lambda}{D}} \left. \frac{d m(\alpha_{\frac{\lambda}{D}})}{d \alpha_{\frac{\lambda}{D}}} \right|_{\chi \beta_{\frac{\lambda}{D}} = 0} + \frac{1}{2} \left( \chi \beta_{\frac{\lambda}{D}} \right)^{2} \left. \frac{d^{2} m(\alpha_{\frac{\lambda}{D}})}{d \alpha_{\frac{\lambda}{D}}^{2}} \right|_{\chi \beta_{\frac{\lambda}{D}} = 0}  \nonumber \\
	&=&  m(\alpha_{\frac{\lambda}{D}}) + \frac{w_{0}}{\xi_{x}} \left( \frac{w_{0}\pi \beta_{\frac{\lambda}{D}}}{R} \right) \left\{ \frac{\cos\left(w_{0}\pi \alpha_{\frac{\lambda}{D}}\right)}{w_{0}\pi \alpha_{\frac{\lambda}{D}}} - \frac{\mathrm{sinc}\left(w_{0}\pi \alpha_{\frac{\lambda}{D}}\right)}{w_{0}\pi \alpha_{\frac{\lambda}{D}}} \right\} \nonumber \\
	&\quad & + \frac{w_{0}}{2 \xi_{x}} \left( \frac{w_{0}\pi \beta_{\frac{\lambda}{D}}}{R} \right)^{2}\left \{ - \frac{\sin\left(w_{0}\pi \alpha_{\frac{\lambda}{D}}\right)}{w_{0}\pi \alpha_{\frac{\lambda}{D}}} - 2 \frac{\cos\left(w_{0}\pi \alpha_{\frac{\lambda}{D}}\right)}{\left( w_{0}\pi \alpha_{\frac{\lambda}{D}} \right)^{2}} + 2 \frac{ \mathrm{sinc}\left(w_{0}\pi \alpha_{\frac{\lambda}{D}}\right)}{\left( w_{0}\pi \alpha_{\frac{\lambda}{D}} \right)^{3}}\right \}.
\end{eqnarray}
Given that the light passes through the center of the mask with a length of $2B_{\frac{\lambda{c}}{D}}$, the Fourier conjugate of the third term in the right-hand side of Equation \ref{m_spe_approx_appendix} is
\begin{eqnarray}
	\Delta \tilde{m}_{spectrum,2nd}(x,y) &=& \frac{w_{0}}{2 \xi_{x}} \left( \frac{w_{0}\pi}{R} \right)^{2} \left( \frac{\lambda}{D} \right) \int_{-B_{\frac{\lambda_{c}}{D}}}^{B_{\frac{\lambda}{D}}} d\beta_{\frac{\lambda}{D}} \beta_{\frac{\lambda}{D}}^{2} \mathrm{e}^{-2 \pi i \beta_{\frac{\lambda}{D}} \frac{y}{D}} \nonumber \\ 
	&\quad & \times \int_{-\infty}^{\infty} d\alpha_{\frac{\lambda}{D}} \left \{ - \frac{\sin\left(w_{0}\pi \alpha_{\frac{\lambda}{D}}\right)}{w_{0}\pi \alpha_{\frac{\lambda}{D}}} - 2 \frac{\cos\left(w_{0}\pi \alpha_{\frac{\lambda}{D}}\right)}{\left( w_{0}\pi \alpha_{\frac{\lambda}{D}} \right)^{2}} + 2 \frac{ \mathrm{sinc}\left(w_{0}\pi \alpha_{\frac{\lambda}{D}}\right)}{\left( w_{0}\pi \alpha_{\frac{\lambda}{D}} \right)^{3}} \right \} \mathrm{e}^{-2 \pi i \alpha_{\frac{\lambda}{D}} \frac{x}{D}} \nonumber \\
	&=& \frac{w_{0}}{2 \xi_{x}} \left( \frac{w_{0}\pi}{R} \right)^{2} U (x) \left \{\frac{2B^{2}\sin\left(2\pi B_{\lambda_{c}}\frac{y}{D} \right)}{2\pi \frac{y}{D}} + \frac{4B_{\lambda_{c}} \cos\left( 2\pi B_{\lambda_{c}}\frac{y}{D} \right) }{\left( 2\pi \frac{y}{D} \right)^{2}} - \frac{4 \sin\left( 2\pi B_{\lambda_{c}}\frac{y}{D} \right) }{\left( 2\pi \frac{y}{D} \right)^{3}} \right \} ,
\end{eqnarray}
where $U(x)$ represents the Fourier conjugate of the $\alpha$ component of the third-term in Equation \ref{m_spe_approx_appendix}. $\Delta \tilde{m}_{spectrum,2nd}(x,y)$ rapidly increases around $y=0$, and the $y$ component approaches approximately $B_{\frac{\lambda_{c}}{D}}^{3}$; $\Delta \tilde{m}_{spectrum,2nd}(x,y=0)$ is proportional to $R^{-2}B_{\frac{\lambda_{c}}{D}}^{3}$. When the spectral resolution is fully larger than the length of the mask, $\Delta \tilde{m}_{spectrum,2nd}(x,y=0)$ is rapidly decreasing as the $n$-th order of the Taylor series is higher. Therefore, the first-order Taylor expansion is used for the calculation of the stellar leak in Section \ref{subsec:mask}. 

\section{In-plane non-uniformity of the grating efficiency}
\label{sec:appendix_b}
\begin{figure}
	 \centering
	\includegraphics[scale=0.6,height=5cm,clip]{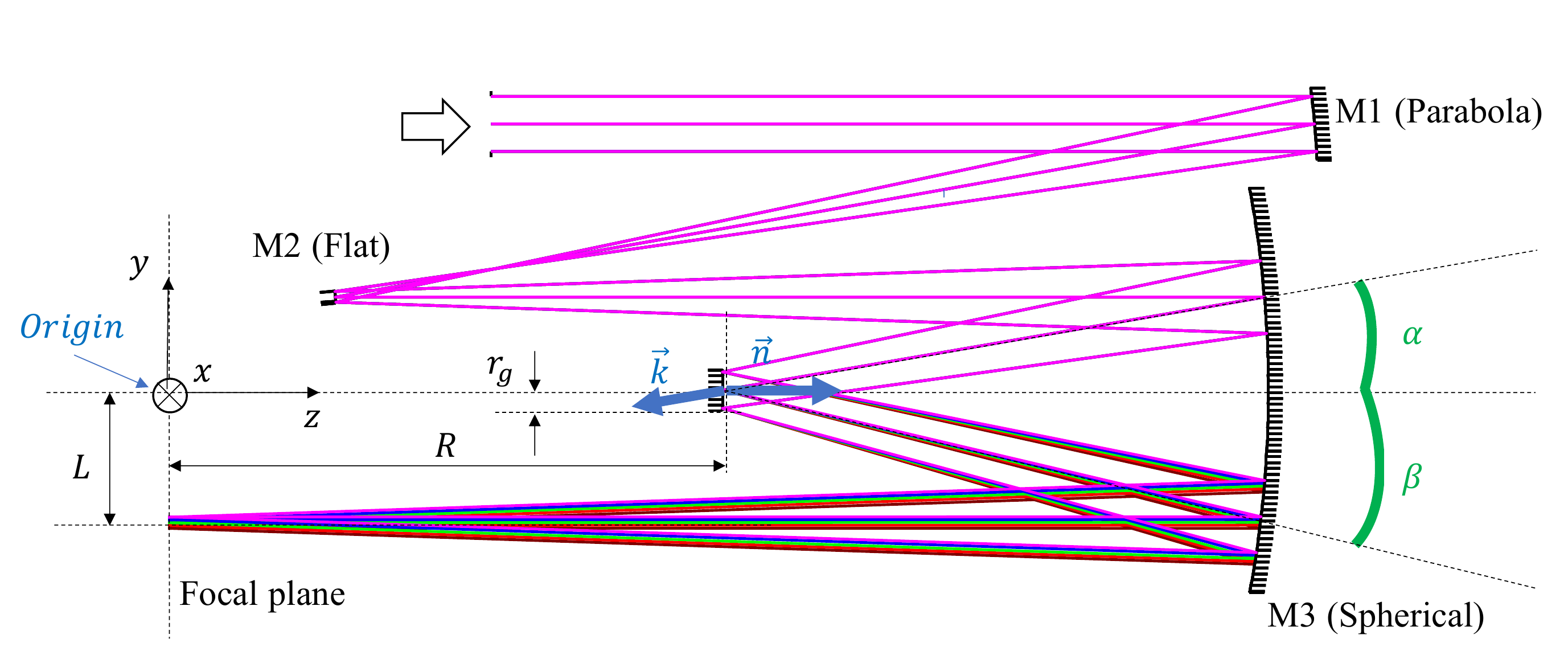}
	\caption{Coordinate system and the parameters for the Offner-type spectrograph.}
	\label{fig:coordinate_system_for_offner}
\end{figure}

We analytically describe in-plane non-uniformity of the grating efficiency, assuming that the convex reflection grating and the spherical mirror are ideally arranged; the aberration is negligible (see Section \ref{subsec:performance}).  Figure \ref{fig:coordinate_system_for_offner} shows the coordinate system and parameters of the Offner-type spectrograph prepared for this analysis. The spectrum is formed along the $y$ axis, as shown in Figure \ref{fig:optical_design}. The grating efficiency on the pupil plane, $\eta(x,y)$, is written as
\begin{equation}
	\label{ita_initial}
	\eta (x,y) = \frac{\sin (m \pi (B(x,y) - 1))}{m \pi (B(x,y) -1)},
\end{equation}
where $m$ is the diffraction order, and $B(x,y)$ shows the optical path length between the diffraction grating and the focal plane. Given that the incident and exit angles are set to $\alpha$ and $\beta$, respectively, $B(x,y)$ is 
\begin{equation}
	B (x,y) = \frac{d \tan \epsilon}{\lambda} (\cos \alpha(x,y) + \cos \beta(x,y)),
\end{equation}
where $\epsilon$ is the blaze angle of the diffraction grating. In addition, $\cos \alpha$ can be described as the inner product of the normal vector of the diffraction grating, $\vec{n}$, and the unit vectors of the incident beam, $\vec{k}$, which are shown in Figure \ref{fig:coordinate_system_for_offner}. The two unit vectors, $\vec{n}$ and $\vec{k}$, are expressed as
\begin{eqnarray}
	\vec{n} &=& \frac{1}{r} (x, y, \sqrt{r^{2} - x^{2} + y^{2}}) \nonumber \\
	\vec{k} &=& \frac{1}{\sqrt{r^{2} + 2 y l + l^{2}}} (x, y+r_{g}l, z), 
\end{eqnarray}
where $l$ and $r$ are $L$ and $R$ divided by the radius of the diffraction grating, $r_{g}$, respectively; $L = l r_{g}$ and $R = r r_{g}$. $\cos \alpha$ is expressed using the coordinate system, $(x,y)$: 
\begin{eqnarray}
	\cos \alpha (x,y) &=& \frac{1 + \frac{l y}{r^{2}}}{\sqrt{1 + \frac{l^{2} + 2 l y}{r^{2}}}} \nonumber \\
		&\simeq & 1 - \frac{l^{2}+l y}{r^{2}}.
\end{eqnarray}
We approximated the above equation, assuming that $r$ is much larger than $l$ and $y$.  

The incident and exit angles are related through the following equation: 
\begin{equation}
	\label{diff_order}
	m \lambda = d (\sin \alpha(x,y) - \sin \beta(x,y)) \cos \gamma (x,y),
\end{equation}
where $d$ shows the pitch of the diffraction grating, and $\gamma$ is the incident angle to the grating around the $y$ axis. Therefore, $\cos \gamma(x,y)$ is written as follows:
\begin{eqnarray}
	\label{cos_gamma}
	\cos \gamma (x,y) &=& \frac{\sqrt{1 + \frac{l^{2} + 2 ly - x^{2} -y^{2}}{r^{2}}}}{\sqrt{1 + \frac{l^{2} + 2 l y}{r^{2}}}} \nonumber \\
		&\simeq & 1 - \frac{1}{2}\frac{x^{2} + y^{2}}{r^{2}}.
\end{eqnarray}
Using Equations \ref{diff_order} and \ref{cos_gamma}, we describe $\cos \beta (x,y)$ as  
\begin{eqnarray}
	\cos \beta (x,y) &=& \cos \left \{ \sin^{-1}(- \frac{m \lambda}{ d \cos \gamma} + \sin \alpha) \right \}  \nonumber \\
		&\simeq & 1 - \frac{1}{2} \left( \frac{m \lambda}{d} -  \frac{l}{r} \right)^{2}.
\end{eqnarray}
Based on the above considerations, $B (x,y)$ is: 
\begin{equation}
	B (x,y) \simeq \frac{d \tan \epsilon}{\lambda} \left \{ 2 - \frac{1}{2} \left( \frac{m \lambda}{d} \right)^{2} + \frac{m \lambda}{d} \frac{l}{r} - \frac{3}{2} \left( \frac{l}{r} \right)^{2} - \frac{ly}{r^{2}} \right \}.
\end{equation}
The first term of the right-hand equation is much larger than the other terms, and the sum of all the terms, except for the first term, is replaced with $\sigma (y)$, highlighting the linear dependence of $y$. Through the MacLaughlin expansion, the in-plane diffraction grating shown in Equation \ref{ita_initial} is approximated as follows: 
\begin{eqnarray}
	\label{ita_later}
	\eta (x,y) &\simeq & 1 - \frac{1}{6} m^{2} \pi^{2}  \left( \frac{B(x,y)}{m} -1 \right)^{2}.  \nonumber \\
	&=& 1 -  \frac{1}{6} m^{2} \pi^{2} \left \{ \left( 2 \frac{d \tan \epsilon}{m \lambda} - 1\right) + \frac{d \tan \epsilon}{m \lambda} \sigma (y) \right\}^{2} \nonumber \\
	&\simeq & 1 -  \frac{1}{6} m^{2} \pi^{2} \left \{ \left( 2 \frac{d \tan \epsilon}{m \lambda} - 1\right)^{2} + 2 \left( 2 \frac{d \tan \epsilon}{m \lambda} - 1\right) \sigma (y) \right \}.
\end{eqnarray}
Thus, the in-plane diffraction efficiency is written as the linear dependence of the spectral direction. 

\bibliography{LUVOIR_coronagraph}{}
\bibliographystyle{aasjournal}


\end{document}